\documentclass[sigconf,a4paper]{acmart}
\usepackage[T1]{fontenc}
\usepackage{hyperref}
\usepackage{textcomp}
\usepackage{amsmath}
\usepackage{subcaption}
\usepackage{xcolor}
\usepackage{booktabs, enumitem, graphicx, multirow, cleveref}
\usepackage{makecell}
\usepackage{siunitx}
\usepackage{comment, amsfonts}
\usepackage{amsthm}
\usepackage{multicol}
\usepackage{tikz}
\usetikzlibrary{shapes.geometric}
\usepackage[vlined, linesnumbered]{algorithm2e}
\SetAlCapSkip{0pt}

\newcommand{\legtri}{\raisebox{0.15ex}{\large$\blacktriangle$}}
\newcommand{\legsquare}{\raisebox{0ex}{$\blacksquare$}}
\newcommand{\legcirc}{\raisebox{-0.2ex}{\Large$\bullet$}}

\usepackage{amsmath}

\makeatletter
\renewcommand\p@subfigure{\thefigure} 

\makeatother

\AtBeginDocument{%
  }

\setcopyright{acmlicensed}
\copyrightyear{2027}
\acmYear{2027}
\acmDOI{XXXXXXX.XXXXXXX}
\acmConference[EDBT '27]{30th International Conference on Extending Database Technology}{April 06--09,
  2027}{Lille, France}
\acmISBN{978-1-4503-XXXX-X/2018/06}

\newcommand{\ours}{U-HNSW}

\setlist{itemsep=2pt, topsep=3pt, parsep=2pt, partopsep=2pt, leftmargin=*}

\begin{document}
\title{\ours: An Efficient Graph-based Solution to ANNS Under Universal $L_p$ Metrics}
\author{
	Huayi Wang \qquad
	Jingfan Meng \qquad
	Jun Xu
}

\affiliation{%
	\institution{%
		\ttfamily
		\{hwang762, jmeng40\}@gatech.edu, jx@cc.gatech.edu
	}
	\institution{%
		\normalfont
		Georgia Institute of Technology, Atlanta, USA
	}
	\country{}
}

\authorsaddresses{}

\renewcommand{\shortauthors}{Wang, Meng, and Xu}
\begin{abstract}

Approximate nearest neighbor search under universal $L_p$ metrics 
(ANNS-U-$L_p$) is an important and challenging research problem, as it requires answering queries under all possible $p$ ($0.5<p \le 2$) values simultaneously without building an index for each possible $p$ value. 
The state-of-the-art solution, called MLSH, is a Locality-Sensitive Hashing (LSH)-based ANNS method with barely acceptable query performance. 
In contrast, graph-based ANNS methods, which offer significantly improved query efficiency on the ANNS-$L_p$ problem (with a fixed $p$-value), cannot be naively extended to the ANNS-U-$L_p$ problem. 
In this paper, we propose U-HNSW, the first graph-based method for ANNS-U-$L_p$. 
Our scheme uses HNSW graph indexes built on two base metrics ($L_1$ and $L_2$) to generate promising NN candidates, 
and then verifies these candidates with an early-termination strategy. 
Experimental results show that U-HNSW not only achieves up to 2670 times shorter query time than the original MLSH implementation running on a RAM disk, but also outperforms the original HNSW on the ANNS-$L_p$ problem, except for a few special $p$ values.

\end{abstract}
\maketitle
\section{Introduction}\label{sec:intro}
Approximate nearest neighbor search (ANNS)
is a fundamental algorithmic problem arising in many areas of computer science,
such as large language models~\cite{ann_llm1,ann_llm_app2}, computer vision~\cite{ann_cv},
and information retrieval~\cite{Lin_retrieval_2015}.
In these applications, data items in a dataset $\mathcal{D}$ are represented by vectors in a 
high-dimensional (say $d$-dimensional) space.  Given a query point $\vec{q}$,
the goal of ANNS is to find one or more points in $\mathcal{D}$ that are among the closest
to $\vec{q}$ according to a certain distance metric, such as $L_2$ distance.

\subsection{ANNS-U-$L_p$: Application and Challenges}\label{subsec:annsu_prob}
In this work, we attack a fundamentally different variant of the ANNS
problem. 
In this problem, we need to
answer queries over a dataset $\mathcal{D}$ that lies in a $d$-dimensional space.
However, unlike traditional ANNS formulations that assume a fixed distance metric, each query in our problem specifies its own distance metric parameter $p$.
Specifically, the $i$-th query is a tuple $(\vec{q}_i, p_i)$, where $\vec{q}_i$ is the query vector and $p_i$ determines the $L_p$ metric used to measure distances.
The objective is to find $\vec{x} \in \mathcal{D}$ that is closest to $\vec{q}_i$ under the $L_{p_i}$ distance.
Because the value of $p_i$ can vary from one query to the next, we refer to this problem as ANNS under universal $L_p$ metrics, or ANNS-U-$L_p$ for short.



While the $L_2$ metric is undoubtedly the most widely used and influential 
distance measure in ANNS~\cite{hnsw,nsg,ann_benchmark,hvs_graph}, other 
$L_p$ metrics are also important, as they can often yield more insightful 
and discriminative results in data mining, image retrieval, and genomic 
data search~\cite{lazylsh,image_p,gene_search}.
Accordingly, the ANNS problem under a general $L_p$ metric (ANNS-$L_p$) 
has been studied extensively in the literature~\cite{p_lsh,p_qalsh}.
However, an ANNS-$L_p$ query engine built for a fixed value of $p$ is 
often inadequate for many modern ML applications, where the most 
discriminative choice of $p$ can vary across datasets or tasks and must 
be tuned per application~\cite{lazylsh}. For example,
	in $K$-NN classification, \cite{lazylsh} shows that the optimal $p$
	differs across datasets: $p = 0.5$ achieves the highest accuracy on
	a physical signal dataset, $p = 0.6$ is optimal on an image dataset,
	while $p = 1.0$ is optimal on a text dataset. Since the best $p$ cannot
	be known in advance and must be determined empirically, an efficient
	query engine that supports arbitrary $L_p$ metrics without rebuilding
	the index is essential. An efficient query engine for ANNS
under universal $L_p$ metrics (ANNS-U-$L_p$) would be ideal for such 
applications, as it enables adaptive exploration of the dataset 
$\mathcal{D}$ under different $L_p$ metrics---allowing each application 
to identify the most suitable, task-specific $p$ value dynamically.

The ANNS-U-$L_p$ problem is particularly challenging because it demands high query efficiency and accuracy across a continuum of $L_p$ metrics.
A straightforward approach is to construct, for each possible $p_i$ value, a dedicated ANNS index (e.g., using $p$-stable Locality-Sensitive Hashing~\cite{p_lsh}) to handle ANNS-$L_{p_i}$ queries.
However, this naive solution is impractical, as it would require building and maintaining a large number of separate indexes for the full range of $p$ values in $[0.5, 2]$, even after discretization. For example, a coarse discretization of $p$ at intervals of 0.1 (as used in prior work~\cite{lazylsh, mlsh}) would require
16 separate indexes.

\subsection{Existing Solutions}\label{subsec:intro_mlsh}

LazyLSH~\cite{lazylsh} represents the first practical attempt to address the ANNS-U-$L_p$ problem without building multiple indexes.
It is built upon the $p$-stable LSH framework~\cite{p_lsh} and aims to reuse a single index constructed for the $L_1$ metric (called the {\it base metric} in~\cite{lazylsh}) to answer queries under various $L_p$ metrics for $0.5 \leq p \leq 1$.
The key insight is that two points $\vec{x}$ and $\vec{y}$ that are close under the $L_1$ metric are also not too far under the $L_p$ metric, provided that $p$ is not too far from~1.
However, as $p$ deviates further from~1 (e.g., toward~0.5), this correlation between $L_1$-closeness and $L_p$-closeness diminishes, leading to a noticeable degradation in query accuracy, as observed in~\cite{mlsh}.

Multi-Metric LSH (MLSH)~\cite{mlsh}, the state-of-the-art solution for the ANNS-U-$L_p$ problem, addresses this limitation by introducing an additional LSH index for the $L_{0.5}$ metric “at the other end.”
Specifically, given an ANNS-U-$L_p$ query $(\vec{q}_i, p_i)$, MLSH selects which index to use based on the value of $p_i$: when $p_i$ is closer to 0.5 than to~1, it relies on the $L_{0.5}$ index instead of the $L_1$ index to find the nearest neighbors 
(NNs) of~$\vec{q}_i$.
However, the overall query performance of MLSH remains constrained by the inherent inefficiency of the LSH framework, which is known to incur relatively long query times in practice~\cite{hnsw,ann_benchmark}.

\subsection{Our Graph-based Solution to ANNS-U-$L_p$}\label{subsec:intro_graph}

For ANNS-$L_p$ with a fixed $p$, graph-based methods such as Hierarchical Navigable Small World (HNSW)~\cite{hnsw} have been shown to deliver substantially better query efficiency
than LSH-based approaches.
As a result, graph-based ANNS methods have become one of the most widely adopted approaches in modern industrial vector database systems~\cite{Pinecone,Milvus}.
We are hence motivated to explore whether a graph-based ANNS-$L_p$ solution can be extended to handle the ANNS-U-$L_p$ problem.
However, a naive approach\textemdash constructing a separate graph index for every possible $p$ value\textemdash clearly fails to scale.
Moreover, building even a single high-quality graph index is computationally expensive and time-consuming, as noted in~\cite{ann_benchmark}.
These challenges likely explain why, despite the appeal of graph-based methods, no prior work has successfully extended them to support efficient ANNS under universal $L_p$ metrics\textemdash until this work. 


Another challenge posed by ANNS-U-$L_p$ is that, even for the much simpler ANNS-$L_p$ problem 
(with a fixed $p$), directly applying HNSW is much less efficient (except for a few special $p$ values) than its well-known performance for ANNS-$L_2$. This is because the dominant cost in HNSW query processing arises from query-to-data (Q2D) distance computations, and the algorithm needs to evaluate Q2D distances for a large number of intermediate points.
When the target metric is a general $L_p$, each such Q2D computation is substantially more expensive than under $L_2$ (as analyzed in Section~\ref{subsec:anns_lp_time}), which can significantly increase the query time.

To address the above challenges, we propose Universal HNSW 
(U-HNSW)\textemdash the first graph-based method that extends the 
widely adopted HNSW algorithm for solving the ANNS-U-$L_p$ problem.
Unlike a naive approach that simply replaces the LSH
	indexes in MLSH with HNSW graphs, U-HNSW is enabled by two key
	observations: 1) $L_1$ or $L_2$ distance is over an order
	of magnitude faster to compute than general $L_p$ via SIMD as described in Section~\ref{subsec:anns_lp_time}; 2) $L_1$ or $L_2$
	metric ordering is a highly effective filter for $L_p$ nearest
	neighbors, so a small candidate set of only $t = 300$ points already captures the true top-$K$ $L_p$ NNs with recall close to 1, as we detail in Section~\ref{subsec:algodesc}.
	Specifically, 
	U-HNSW maintains only two graph
	indexes: $G_1$, an HNSW index built under the $L_1$ metric, and $G_2$,
	an HNSW index built under the $L_2$ metric.
During query processing, given a query $(\vec{q}, p)$, U-HNSW first 
selects the base index\textemdash either $G_1$ or $G_2$\textemdash 
whose underlying metric is closer to the query metric $L_p$ (with a 
slight weighting adjustment). It then searches for ANNs of $\vec{q}$, in the selected base metric ($L_1$ or $L_2$), 
using the selected graph ($G_1$ or $G_2$).  This step is 
computationally efficient because 
Q2D distance computations 
in the base metric ($L_1$ 
or $L_2$)  is relatively inexpensive 
(observation 1). Next, each candidate point 
returned from the base-metric search\textemdash ordered by its base 
distance (e.g., $L_1$)\textemdash is vetted by computing its exact Q2D 
distance under the query metric $L_p$. This step is also computationally efficient thanks to observation 2.
To further improve query efficiency, U-HNSW performs these relatively expensive $L_p$
distance computations progressively and terminates once the estimated recall is found to
exceed the target threshold as described in Section~\ref{subsec:algodesc}.
As 
demonstrated in Section~\ref{sec:eval}, U-HNSW not only achieves a 
substantial performance improvement over MLSH, but also significantly 
outperforms the original HNSW when applied to fixed-metric 
ANNS-$L_p$ problems, except for a few special $p$ values (e.g., $p=0.5, 1.5$).

In summary, this work makes three main contributions.  First, we present U-HNSW, the first cost-effective practical graph-based solution to the ANNS-U-$L_p$ problem. 
Second, we develop an adaptive candidate checking framework that reduces the number of expensive $L_p$ distance computations, allowing U-HNSW to outperform its baseline HNSW even for fixed-metric ANNS-$L_p$ tasks.
Third, through extensive experiments, we show that  U-HNSW not only achieves up to 2670 times shorter query times than the original MLSH implementation running on a RAM disk (up to 15 times shorter than the idealized MLSH), but also outperforms the original HNSW on the ANNS-$L_p$ problem,except for a few special $p$ values.
\section{Background and Related Work}\label{sec:background}

In this section,
we first define the problem of ANNS under universal $L_p$ metrics (ANNS-U-$L_p$) and explain how the computational cost of $L_p$ distance varies significantly with $p$, in Section~\ref {subsec:anns_lp_time}.  
Finally, we describe HNSW, which serves as the baseline for our U-HNSW scheme, in Section~\ref{subsec:hnsw_background}.

\subsection{ANNS under Universal $L_p$ Metrics}\label{subsec:anns_lp_time}


Given a query tuple $(\vec{q},p)$, the goal of ANNS-U-$L_p$ is to retrieve the top-$K$ nearest neighbors of the query point $\vec{q}$ in $\mathcal{D}$ according to the $L_p$ metric at a high recall value (say 90\%) in as little query time as possible. In the sequel, unless specifically mentioned, we assume that the given $p$ value ranges from 0.5 to 2, which covers the commonly evaluated range of $p$ values used in previous ANNS-U-$L_p$ works~\cite{lazylsh,mlsh}. We follow the convention adopted in the ANNS literature~\cite{hnsw,mlsh,ann_benchmark,p_qalsh,lazylsh,opq,rabitq} of focusing on $p \leq 2$, since $L_p$ metrics with $p > 2$ are rarely used in practical ANNS applications. When the dimension $d$ is not small, the computation time of this $L_p$ distance can vary significantly for different values of $p$, as follows.



\begin{figure}[!htb]
   \centering
 \begin{minipage}{\linewidth}
	\centering
	\scriptsize
	\setlength{\tabcolsep}{1.5pt}
	\renewcommand{\arraystretch}{0.9}

 	\begin{tabular}{@{}l@{\hspace{0.35em}}l@{\hspace{0.9em}}l@{\hspace{0.35em}}l@{\hspace{0.9em}}l@{\hspace{0.35em}}l@{}}
 		\textcolor{blue}{\rule[0.8ex]{0.60cm}{1.1pt}}%
 		\hspace{-0.33cm}\textcolor{blue}{\legtri}\hspace{0.18cm} & \textcolor{blue}{General $L_p$} &
 		\textcolor{red}{\rule[0.8ex]{0.60cm}{1.1pt}}%
 		\hspace{-0.33cm}\textcolor{red}{\legsquare}\hspace{0.18cm} & $L_{0.5}/L_{1.5}$ &
 		\textcolor{violet}{\rule[0.8ex]{0.60cm}{1.1pt}}%
 		\hspace{-0.33cm}\textcolor{violet}{\legcirc}\hspace{0.18cm} & $L_1/L_2$
 	\end{tabular}
 	\vspace{0.10cm}
 \end{minipage}

   \centering
   \includegraphics[scale=0.21]{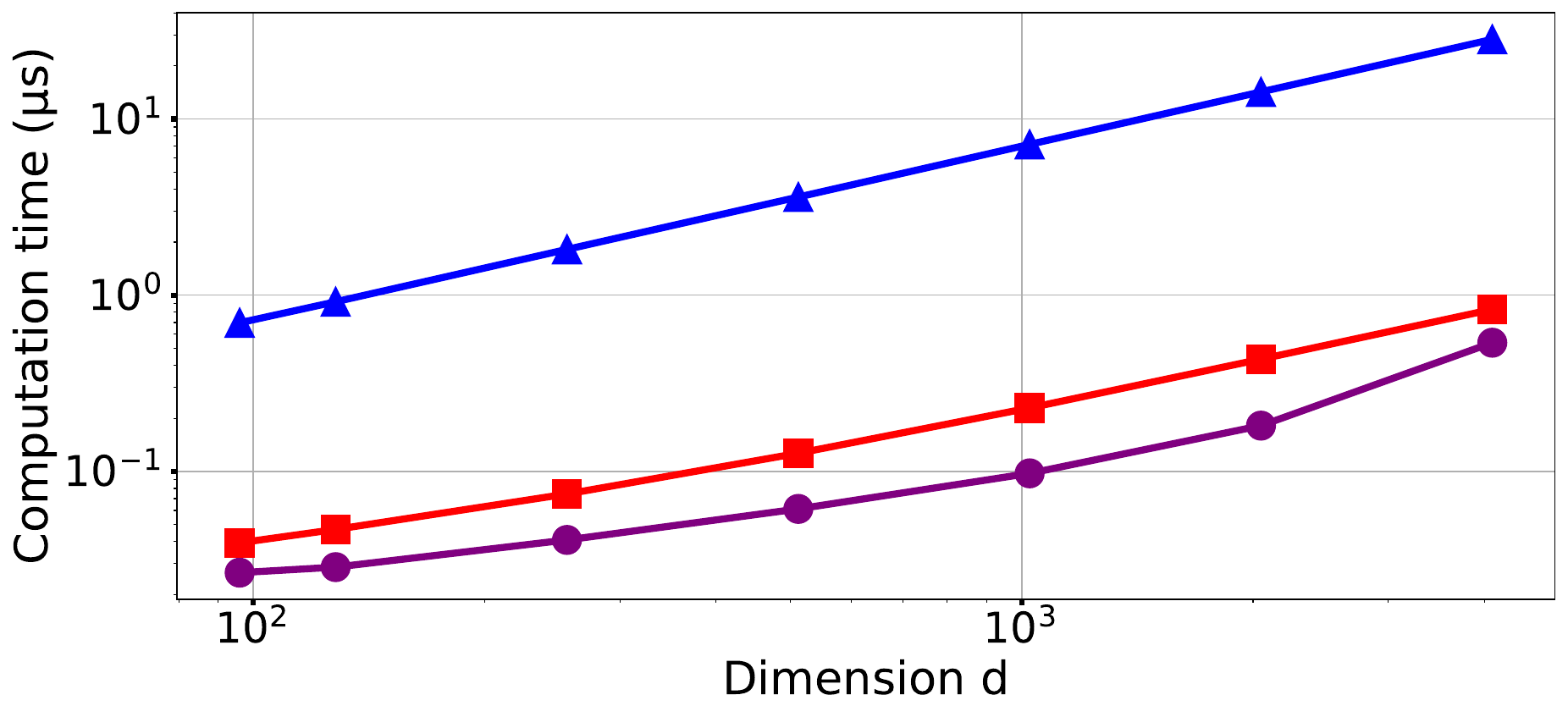}
   \caption{
     Average computation times (in microseconds) of each $L_p$ distance instance on our
     workstation described in Section~\ref{sec:eval-setup} (both axes in log scale).
     $L_1$ and $L_2$ metric computations are faster than general $L_p$ metric computation by more than one
     order of magnitude.
   }\label{fig:computation_time}
 \end{figure}

\begin{itemize}
\item \textbf{$p=1$ or $2$ (fastest).}  The computation of the $L_1$ or $L_2$ metric involves only basic arithmetic operations (add/subtract/multi-ply), which can be performed in a computationally efficient manner using SIMD instructions such as \texttt{\_mm512\_sub\_ps} and \texttt{\_mm512\_add\_ps}. As shown in Figure~\ref{fig:computation_time}, the $L_1$ or $L_2$ metric is the fastest to compute across different dimensionalities.

\item \textbf{$p=0.5$ or $1.5$ (quite fast).}  In addition to basic arithmetic operations, the computation of the $L_{0.5}$ or $L_{1.5}$ metric involves the square root operation, which can also be computed efficiently with SIMD instructions such as
\texttt{\_mm512\_sqrt\_ps}. As shown in Figure~\ref{fig:computation_time}, the computation of the $L_{0.5}$ or $L_{1.5}$ metric is slightly slower than that of the $L_1$ or $L_2$ metric across different $d$ values.

\item \textbf{Other $p$ values (slow).} For $L_p$ metrics with other $p$ values,
power operations are needed in their computations, which are not SIMD-friendly
and are much more expensive computationally than the aforementioned basic and
square root operations. Since all such $p$ values require
power operations, their computation times are nearly identical to each other.
As a result, computation time of a general $L_p$ metric is more than an order
of magnitude slower than that of the $L_1$ or $L_2$ metric as shown in Figure~\ref{fig:computation_time}.
\end{itemize}

\subsection{Hierarchical Navigable Small World Graph} \label{subsec:hnsw_background}
Hierarchical Navigable Small World (HNSW) graph~\cite{hnsw} is an 
efficient and widely recognized graph-based solution to ANNS-$L_p$ 
for a fixed $p$ value ($0.5 \leq p \leq 2$). HNSW organizes the 
dataset into a multi-layer proximity graph, where each node represents 
a data point. The graph is built incrementally by
inserting data points one by one: each new point is connected to its
$M$ closest neighbors in the graph (where $M$ is a key construction
parameter), and the quality of these connections is controlled by
$\texttt{efConstruction}$, the size of the dynamic candidate list
maintained during insertion. Larger $M$ and $\texttt{efConstruction}$
yield higher-quality graphs at the cost of longer construction time.
If two nodes are close to each other in the $L_p$ metric, there is 
an edge connecting them in the proximity graph. Given a query point 
$\vec{q}$, HNSW uses the following greedy search strategy to find 
the top-$K$ NNs: starting from a pre-defined entry point, HNSW 
traverses the proximity graph along these edges. When reaching a 
previously unvisited node $\vec{v}$, HNSW adds $\vec{v}$ to the NN 
candidate set $\mathcal{C}$, if $\vec{v}$ has a smaller Q2D distance 
than the farthest point in $\mathcal{C}$. The size
of $\mathcal{C}$ is controlled by $\texttt{efSearch}$: a larger
$\texttt{efSearch}$ leads to higher recall at the cost of longer query
time. The top-$K$ NNs are then returned as the $K$ points in
$\mathcal{C}$ with the smallest Q2D $L_p$ distances to $\vec{q}$.




Now we explain why conventional HNSW must build a separate graph index for every possible $p$-value ($L_p$ metric) in order to answer ANNS-U-$L_p$ queries.  As just explained, the HNSW index is constructed by connecting data points (by 
proximity edges) based on their closeness (proximity) under a given $L_p$ metric.  Let $p_1$ and $p_2$ be two distinct $p$ values that are not very close to each other.  Then the set of proximity edges in the graph index built for the $L_{p_1}$
metric can differ significantly from that in the graph index built for the $L_{p_2}$ metric.  Since the ``soul" of a graph index is these proximity edges, 
using the index built for $L_{p_1}$ to search for NNs under $L_{p_2}$ cannot guarantee high query accuracy.  For this reason, HNSW and other graph-based methods~\cite{hnsw,nsg,hvs_graph} all specify a single target distance metric (usually the \(L_2\) metric) and evaluate performance only under that metric.

\subsection{Product Quantization ANNS Methods}\label{subsec:other_anns}

Besides LSH-based and graph-based methods, product
quantization (PQ) based methods~\cite{pq,opq,rabitq} are another widely used
family of ANNS methods. These methods quantize (cluster) all data points in $\mathcal{D}$ 
to a small set of designated points (centroids) and the search is performed over 
only those points in $\mathcal{D}$ that are quantized to the same centroid (and nearby centroids) as the query point is. 
Most of the PQ-based solutions focus exclusively on the ANNS-$L_2$ problem.  
It is not clear how to extend them for solving ANNS-U-$L_p$. 
For example, RaBitQ~\cite{rabitq} does not work for any ANNS-$L_p$ with $p\not=2$, since its conversion of 
$L_2$ to inner product distance does not work for other $L_p$.
A few other PQ methods such as Optimized Product Quantization 
(OPQ)~\cite{opq} could in principle be adapted for ANNS-$L_p$.
However, like graph-based methods, the set of centroids
in a PQ index built for $L_{p_1}$ can differ significantly 
from that built for $L_{p_2}$, when $p_1$ and $p_2$ are not close to 
each other.  
Hence, to answer
ANNS-U-$L_p$ queries, OPQ would  
require building a separate index for each possible $p$ 
value---like that described in
Section~\ref{subsec:annsu_prob}.  For these reasons, we do not compare with 
PQ-based methods in our evaluation.

\section{Scheme Description}
\label{sec:method}

In this section, we present our novel solution \ours~for the ANNS-U-$L_p$ problem.
We first describe the overall procedure of the algorithm, including candidate generation and verification in Section~\ref{subsec:algodesc}.
Then, we explain the rationale behind using graph indexes on the $L_1$ and $L_2$ metrics to generate the candidate set for universal $L_p$ metrics and analyze how to set parameters to minimize the number of expensive $L_p$ distance computations in Section~\ref{subsec:param}. In this section, we focus on the general case where $p \notin \{1,2\}$; when $p=1$ (or $p=2$), the query reduces to standard ANNS under the $L_1$ (or $L_2$) metric, and we can directly search the corresponding graph index $G_1$ (or $G_2$) for $K$-NNs.

\subsection{Algorithm Description}
\label{subsec:algodesc}

As mentioned earlier, U-HNSW extends HNSW for ANNS-U-$L_p$.
To this end, U-HNSW builds and uses two HNSW indexes, namely $G_1$ for the $L_1$ metric and $G_2$ for the $L_2$ metric.
Algorithm~\ref{alg:lpquery} shows how U-HNSW processes an ANNS-U-$L_p$ query $(\vec{q},p)$, to arrive at the $K$ NNs of $\vec{q}$ under the $L_p$ metric.
It consists of the following two steps. 

\begin{figure}[!t]
	\centering
	
	\begin{minipage}{\linewidth}
		\centering
		\scriptsize
		\setlength{\tabcolsep}{3pt}
		\renewcommand{\arraystretch}{0.9}

		\begin{tabular}{@{}l@{\hspace{0.35em}}l@{\hspace{1.0em}}l@{\hspace{0.35em}}l@{}}
			\textcolor{red}{\rule[0.8ex]{0.70cm}{1.2pt}}%
			\hspace{-0.38cm}\textcolor{red}{\legcirc}\hspace{0.20cm} & $G_1$ ( base metric $L_1$ ) &
			\textcolor{blue}{\rule[0.8ex]{0.70cm}{1.2pt}}%
			\hspace{-0.38cm}\textcolor{blue}{\legsquare}\hspace{0.20cm} & $G_2$ (base metric $L_2$)
		\end{tabular}
	\end{minipage}
	
	\begin{subfigure}[b]{0.235\textwidth}
		\centering
		\includegraphics[width=\textwidth]{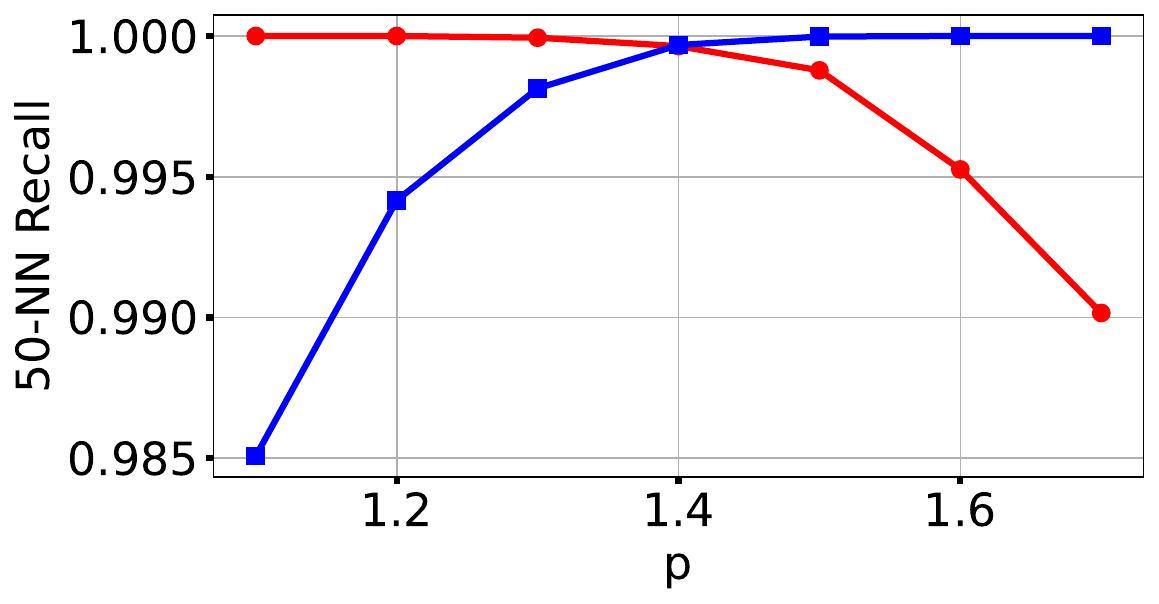}
		\caption{SIFT}
		\label{fig:p0_sift}
	\end{subfigure}
	\hfill
	\begin{subfigure}[b]{0.235\textwidth}
		\centering
		\includegraphics[width=\textwidth]{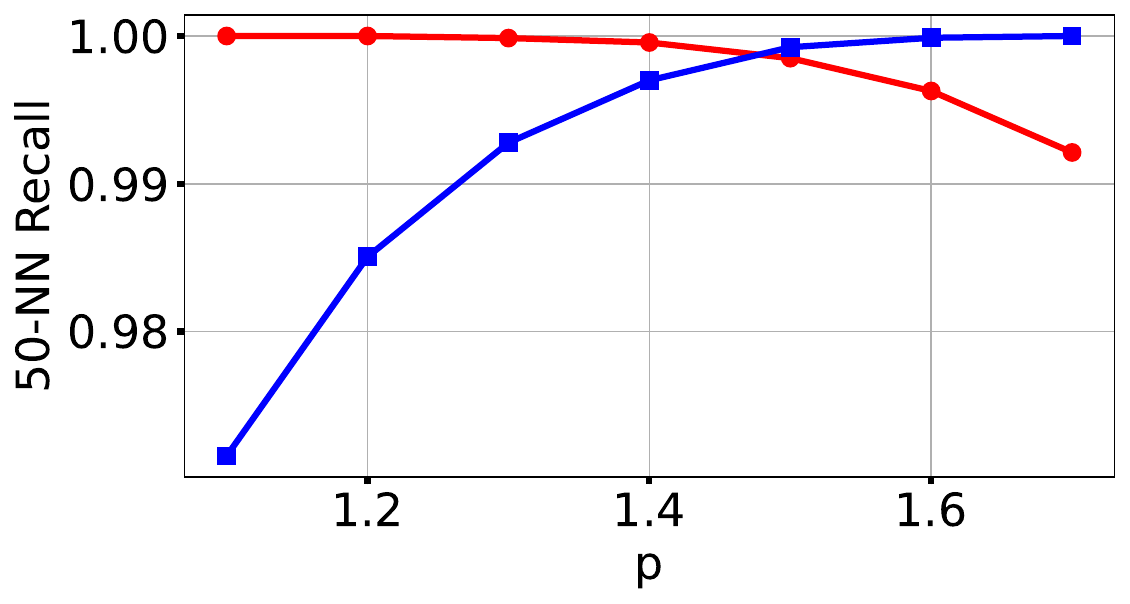}
		\caption{GIST}
		\label{fig:p0_gist}
	\end{subfigure}
	
	\caption{
		 Recall under the target $L_p$ metric when the candidate set is the true top-$t$ NNs on GIST and SIFT datasets in Table~\ref{tab:dataset-stats}.
		We set $t=300$ as described in Section~\ref{subsec:param}.
		The crossover point appears around 1.4, motivating our default cutoff $1.4$.
	}\label{fig:p0}
\end{figure}

\noindent\textbf{Candidate generation.}
U-HNSW first selects a graph index according to the $p$ value specified in the query as follows.
If $p \leq 1.4$, the graph index $G_1$ is selected; otherwise, the graph index $G_2$ is used.
We use $1.4$ as the default cutoff value, which approximately equals the crossover point as shown in Figure~\ref{fig:p0}.
The performance is not sensitive to small perturbations around this value.
Then, U-HNSW runs the standard \texttt{KnnSearch} procedure in HNSW to return a candidate set $C$ containing the $t$-NNs ($t > K$),
sorted in the ascending order of their Q2D distances (as specified in the HNSW code manual~\cite{hnswlib}),
under the corresponding base metric (Line 3 and Line 5).
When $t$ is reasonably large, $C$ empirically contains most of the true top-$K$ NNs of $\vec{q}$ in $L_p$ distance (so a high recall can
potentially be achieved);  we will explain how to set the parameter $t$ in Section~\ref{subsec:param}. This extends the aforementioned observation of LazyLSH (described in Section~\ref{subsec:intro_mlsh}) to a wider range of $p$ values. This extension is justified by the well-known norm equivalence in finite dimensions~\cite{lpnorm_book}.



\noindent\textbf{Candidate verification.}
The candidate set $C$ contains $t$ ($t > K$) points
sorted by $L_1$ or $L_2$ distance, not by the target $L_p$ distance.
Although $C$ contains most of the true top-$K$ NNs (in $L_p$ distance),
simply returning the first $K$ points of $C$ (ranked by their base-metric Q2D distances) may not result in a high recall.
Therefore, we re-rank (verify) points in $C$ by their $L_p$
Q2D distances, as follows.
First, we pop the first $K$ points from the candidate set $C$, and initialize them as (a set) $R$.
Then, we iteratively pop a batch of $\kappa$ ($\kappa > 1$) points from $C$ (denoted as $R'$),
and find the $K$ points (denoted as $R_{new}$) with the smallest $L_p$ Q2D distances in the union $R \cup R'$.
The iteration stops once the intersection ratio \(|R\cap R_{\text{new}}|/K\) is above a pre-defined threshold $\tau$,
indicating that the result will not significantly improve with further iterations.
This early-termination strategy ensures that as few expensive $L_p$ distance computations as needed (to achieve the desired recall) are performed.

The early-termination parameter \(\tau\) can be viewed as an estimator of the recall for the query \(\vec q\).
To make this estimator accurate, the batch size \(\kappa\) must be sufficiently large so that the intersection cardinality \(\lvert R_{\text{new}}\cap R \rvert\) reflects how many candidates in \(R_{\text{new}}\) are likely to
be among the true top-\(K\) NNs of $\vec{q}$.
If \(\kappa\) is too small, U-HNSW may terminate prematurely before examining all potential true top-\(K\) NNs.
We find that setting $\kappa$ to K empirically achieves a good trade-off between query accuracy and query speed. 
\begin{algorithm}[t]
\caption{U-HNSW Query for ANNS-U-$L_p$}\label{alg:lpquery}
\KwIn{Query tuple $(\vec{q},p)$; integer $K$}
\KwOut{Top-$K$ NNs of $\vec{q}$ under $L_p$}
\KwSty{Parameters:} $t$ (candidate set size, default 300 for $K=50$);\quad
$\tau$ (early-termination threshold, default 0.92); \quad\quad\quad\quad\quad\quad
$\kappa$ (batch size, default $K$) \\
\eIf{$p \leq 1.4$}{$C \leftarrow G_1.\texttt{KnnSearch}(\vec{q}, t)$}{$C \leftarrow G_2.\texttt{KnnSearch}(\vec{q}, t)$}
$R \leftarrow$ first $K$ points of $C$\\
\ForEach{batch of $\kappa$ points $R'$ of $C$}{
  $R_{new} \leftarrow$ top-$K$ points by $L_p$ Q2D distance in $R \cup R'$\\
  \lIf{$|R_{new} \cap R|/K \geq \tau$}{\Return $R_{new}$}
  $R \leftarrow R_{new}$
}
\end{algorithm}

\subsection{Parameter Tuning}
\label{subsec:param}

In this subsection, we describe how to tune the parameters in U-HNSW to achieve a short query time while maintaining a good query accuracy (measured by the recall).
As mentioned earlier, in U-HNSW, the total query time $T_{\text{query}}$ consists of two parts: the time for candidate generation ($N_{b}\,T_{b}$) and the time for candidate verification ($N_{p}\,T_{p}$).
That is,
\begin{equation}
T_{\text{query}} \;=\; N_{b}\,T_{b} \;+\; N_{p}\,T_{p},
\label{eq:querycost}
\end{equation}
where $N_{b}$ and $N_{p}$ denote the number of Q2D distance computations under the base metric and under the $L_p$ metric, respectively;
and $T_{b}$ and $T_{p}$ are the average time per Q2D distance computation under the base metric and under the $L_p$ metric, respectively.

In U-HNSW, we primarily tune two parameters: $t$ and $\tau$.
As we will elaborate next, $t$ determines the number of base-metric distance computations $N_b$ in Equation~\ref{eq:querycost},
and $\tau$ determines the number of $L_p$ distance computations $N_p$ in Equation~\ref{eq:querycost}. We now explain how to set these parameters $t$ and $\tau$.

\begin{figure}[t]
    \centering
\begin{minipage}{\linewidth}
  \centering
  \scriptsize
  \setlength{\tabcolsep}{6pt}
  \begin{tabular}{@{}cc@{}}
    \textcolor{green!60!black}{\rule[0.8ex]{0.8cm}{1.2pt}}~GIST&
    \textcolor{red}{\rule[0.8ex]{0.8cm}{1.2pt}}~SIFT
  \end{tabular}

\end{minipage}
    \begin{subfigure}[b]{0.235\textwidth}
        \centering
        \includegraphics[width=\textwidth]{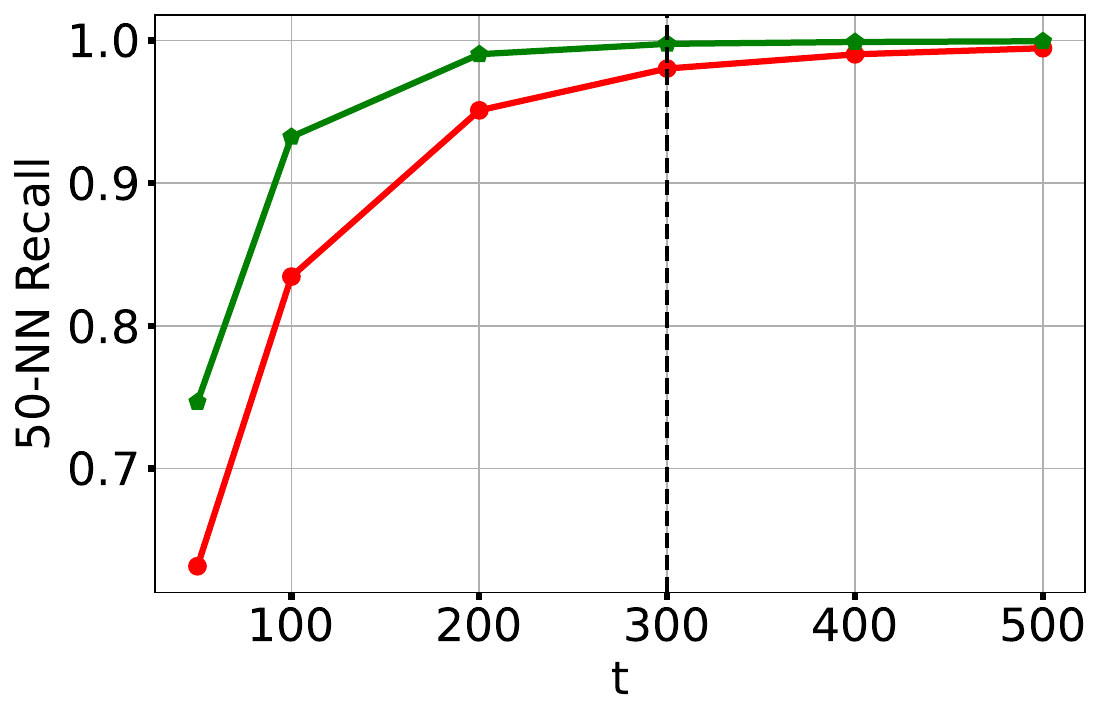}
        \caption{Recall under different $t$.}
        \label{fig:recall_t}
    \end{subfigure}
    \begin{subfigure}[b]{0.235\textwidth}
        \centering
        \includegraphics[width=\textwidth]{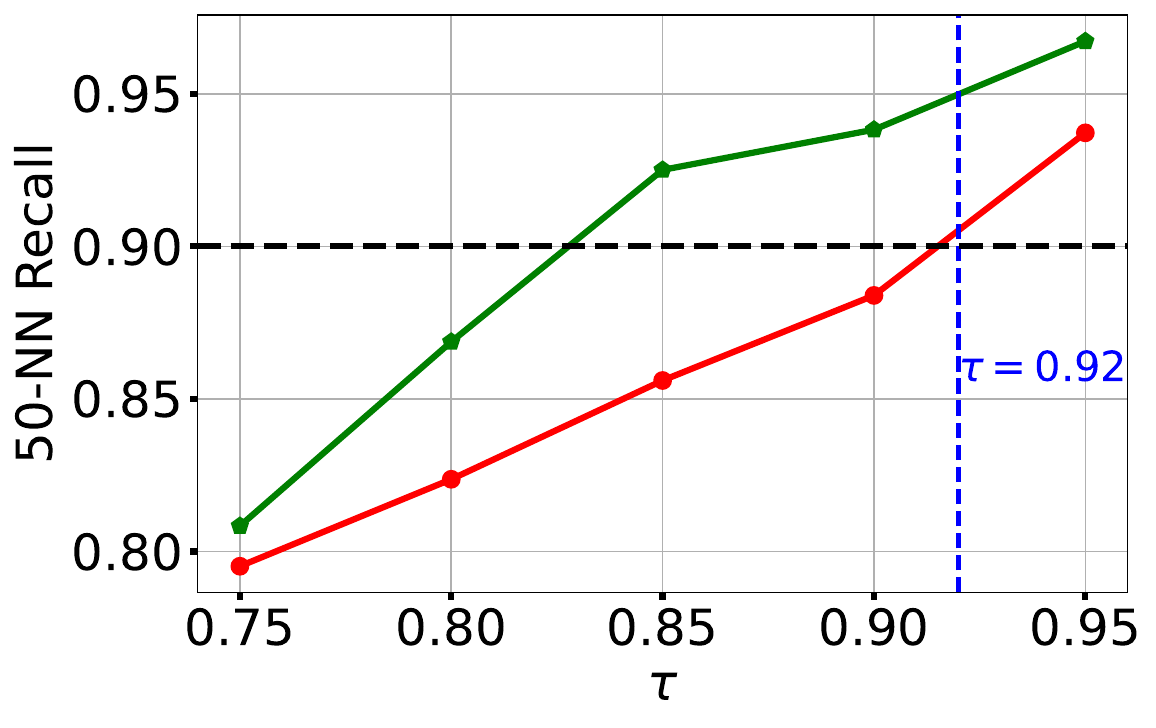}
        \caption{Recall under different $\tau$}
        \label{fig:recall_tau}
    \end{subfigure}
    \caption{Empirical evidence for tuning $t$ and $\tau$ on GIST and SIFT datasets in Table~\ref{tab:dataset-stats}. We use $p=0.5$, which represents the most demanding setting due to its large mismatch to the base metric ($L_1$).}\label{fig:rank}
\end{figure}

As described in Section~\ref{subsec:algodesc}, given $K$, the number of NNs desired by the query, we make $t$ sufficiently large to achieve a target recall value of~$r = 0.9$ (used throughout Section~\ref{sec:eval}).
For $K=50$ (used throughout Section~\ref{sec:eval}), we have found, through the following idealized recall experiment, that $t=300$ is large enough to meet this target recall for all datasets.
Figure~\ref{fig:recall_t} reports the {\it idealized} recall obtained when using the {\it true} top-$t$ NNs under the base metric ($L_1$) as the candidate set.
We evaluate the most demanding setting ($p=0.5$) with $K=50$ on both the SIFT and GIST datasets in Table~\ref{tab:dataset-stats}. These two datasets are chosen as representatives
	because they span a wide range of dimensionalities (128 and 960,
	respectively), and the parameter tuning results on other datasets
	are similar.
As $t$ increases, the idealized recall quickly approaches saturation, so much so that $t=300$ already achieves an idealized recall close to $1$.  In this case, the {\it true potential} recall
(if the Q2D distances of all $t$ points in $\mathcal{C}$ were ``painstakingly'' computed) exceeds 0.94, which leaves a ``wiggle room" of 0.04 for achieving the
actual target recall of 0.9.



Using $t = 300$, which may be larger than necessary
on some datasets, only incurs a small additional overhead in the total
query time, for the following reason.
$N_b$ grows sub-linearly with $t$ when $t$ is within a few hundreds~\cite{hnsw}, and $T_b$ is more than an order of magnitude smaller than $T_p$ (as explained in Section~\ref{subsec:anns_lp_time}). As a result, the term $N_b T_b$ remains small in Equation~\ref{eq:querycost} even if $t$ is larger than necessary.

The threshold parameter $\tau$ controls the early-stopping criterion in U-HNSW's candidate verification step.
As mentioned in Section~\ref{subsec:algodesc}, $\tau$ can be viewed as an estimate of the query recall value.
Given a target recall $r$ (e.g., $r=0.9$ in Section~\ref{sec:eval}), we set $\tau = 0.92$ to leave a small safety margin (a part of the
``wiggle room" above) in the most demanding case, to allow for the recall loss caused by the early-termination strategy.  As shown in Figure~\ref{fig:recall_tau},
setting $\tau = 0.92$ consistently meets the target recall 0.9 on both datasets.
If a higher recall value is required, we can increase $\tau$ accordingly, which allows U-HNSW to meet the new target recall while keeping $N_p$ small, so that the term
$N_p T_p$ remains small in Equation~\ref{eq:querycost}. Similar to the analysis of $t$, we provide empirical evidence from 
two real-world datasets GIST and SIFT in Figure~\ref{fig:recall_tau}. Figure~\ref{fig:recall_tau} clearly shows that setting $\tau=0.92$ (target $0.9$ plus a small margin) consistently meets the target recall (0.9) on both datasets while avoiding unnecessary $L_p$ computations.

\section{Evaluation}\label{sec:eval}
In this section, we conduct an extensive study using six widely used publicly available 
datasets of diverse dimensions, sizes (number of points) and types.
We first evaluate the query performance of U-HNSW and compare it with MLSH's on the ANNS-U-$L_p$ problem in Section~\ref{subsec:comp_mlsh}. Then we evaluate the query performance of U-HNSW on ANNS-$L_p$ (with fixed $p$) 
and compare it with that of HNSW. 
Our results show conclusively that U-HNSW not only outperforms the state-of-the-art ANNS-U-$L_p$ solution MLSH by a factor of up to 15.0 in terms of query efficiency, but also is much faster than its baseline HNSW on ANNS-$L_p$, except for a few special $p$ values.

\subsection{Evaluation Settings}\label{sec:eval-setup}


\subsubsection{Evaluation Datasets.}
We use six widely used publicly available datasets of diverse dimensions, sizes (number of points), and types. The SIFT dataset contains 2 million points sampled uniformly
at random from the 1 billion points contained in SIFT1B~\cite{sift-gist}. 
We
cannot use SIFT1B instead because constructing the HNSW graph index for it would not fit into the \SI{128}{GB} main memory of our workstation.
For each of the
six datasets, Table~\ref{tab:dataset-stats} shows its size $n$ and its dimension
$d$.  For each of the six datasets, the query set contains 1,000 points sampled uniformly at random from the set of query
points associated with the corresponding dataset. 
 
\begin{table}[!htb]
\centering
\small
\setlength{\tabcolsep}{4pt}
\resizebox{\columnwidth}{!}{%
\begin{tabular}{@{}l|c|c|c|c|c|c@{}}
\specialrule{.1em}{.05em}{.05em}
Dataset & Sun~\cite{sun} & Trevi~\cite{Trevi} & GIST~\cite{sift-gist} & Deep~\cite{ann_benchmark} & GloVe~\cite{GloVe} & SIFT~\cite{sift-gist} \\
\specialrule{.1em}{.05em}{.05em}
$n$ & 78K & 99K & 1M & 1M & 1.2M & 2M \\
$d$ & 512 & 4{,}096 & 960 & 256 & 100 & 128 \\
\specialrule{.1em}{.05em}{.05em}
\end{tabular}%
}
\caption{Evaluation datasets summary.}\label{tab:dataset-stats}
\end{table}

\subsubsection{Performance Metrics.}
We evaluate the performance of \ours, HNSW and MLSH in three aspects: index scalability, query efficiency, and query accuracy.
To measure index scalability, we use \textit{index size} (excluding the size of the original dataset $\mathcal{D}$).
To measure query efficiency, we use \textit{query time}. 
To measure query accuracy, we use \textit{recall}, which is defined as follows.
Denote by $\mathcal{S}^*$ the set of true top-$K$ NNs under the $L_p$ metric and by $\mathcal{S}$ the set of $K$ approximate NNs in the result.
The top-$K$ query recall is equal to $|\mathcal{S}^* \cap \mathcal{S}|/K$. Throughout our evaluation, unless otherwise stated, we use $K=50$; using any
other value of $K$ ranging from 10 to 100 (commonly used in the ANNS
literature~\cite{p_qalsh,hnsw,hvs_graph,mlsh}) results in similar query performance. Each query time or recall value presented in this
section is the average over all queries.

\subsubsection{Implementation Details.}\label{subsubsec:imple_details}
We implement \ours~in C++ based on the widely used open-source HNSW implementation called HNSWlib~\cite{hnswlib}. For MLSH, we use the C++ source code provided by its authors.
We optimize the 
$L_1$, $L_2$, $L_{0.5}$ and $L_{1.5}$  distance computations using the SIMD instructions described in Section~\ref{subsec:anns_lp_time} and general $L_p$ distance computations using the optimized implementation in NMSLIB~\cite{nmslib} (a popular ANNS library used in Amazon Elasticsearch service). We compile all C++
source code using g++ 13.3 with -O3.
All experiments are done
on a workstation running Ubuntu 22.04 with an Intel Core
i7\textemdash9800X CPU @ \SI{3.80}{GHz} and \SI{128}{GB} RAM. Our code is available in Anonymous GitHub~\cite{uhnswlib}.

\subsubsection{Benchmark Algorithms.}
In our evaluation, we compare \ours~against the original HNSW and MLSH. We do not compare against LazyLSH since MLSH outperforms LazyLSH as shown in~\cite{mlsh}. The settings of benchmark algorithms are described as follows:

    \textbf{U-HNSW}. For U-HNSW, we first describe the parameter settings of the $G_1$ and $G_2$ graph indexes. For these two graph indexes, we set their parameters so that they can achieve a very high recall when searching for nearest neighbors under the base metric. Specifically, we use $M = 32$ and $\mathit{efConstruction} = 500$ for both $G_1$ and $G_2$. The $\mathit{efSearch}$ parameter is tuned per dataset to ensure that the aforementioned potential recall exceeds 0.94 as described in the third paragraph of Section~\ref{subsec:param}: 400 on Sun, Trevi, GIST and SIFT, 600 on Deep, and 2{,}000 on GloVe.
 Then we set the parameter $\tau =0.92$ which is slightly higher than the target recall value 0.9 as described in Section~\ref{subsec:param}. Such a setting ensures that most queries achieve a recall value larger than 0.9. As for the parameter $t$, we set it to 300 according to the parameter tuning strategy described in Section~\ref{subsec:param}.



    \textbf{HNSW}. 
We compare U-HNSW with the original HNSW on the ANNS-$L_p$ problem with a fixed $p$ value. For $p = 1$ and $p = 2$, the ``original HNSW'' baseline directly reuses $G_1$ and $G_2$, respectively. For other $p$ values, we build a separate HNSW graph index under the corresponding $L_p$ metric using $M = 32$ and $\mathit{efConstruction} = 200$.
The parameter $\mathit{efSearch}$ is tuned per dataset so that the original HNSW achieves an average recall comparable to that of U-HNSW (above 0.9), ensuring a fair comparison of query times at similar recall levels: 200 on Sun, Trevi and SIFT, 280 on GIST, 400 on Deep, and 1{,}600 on GloVe.

    \textbf{MLSH}. We use the parameter settings and the original implementation of MLSH provided by its authors.
    	For our in-memory evaluation (with the MLSH index stored on a RAM disk),
  the measured query time of MLSH is extremely high: For example, on the Sun and GloVe datasets, the original MLSH
  implementation takes approximately 4.67 seconds and 5.35 seconds per query,
  respectively, making U-HNSW 2670 and 890 times faster on these two datasets, respectively.
    	However, one may argue that directly comparing against the original MLSH implementation could be unfair for the following two reasons.
    	First, the $L_p$ distance computations in MLSH are not optimized using the SIMD instructions described in Section~\ref{subsec:anns_lp_time} and the $L_p$ metric optimization in NMSLIB described in Section~\ref{subsubsec:imple_details}, which makes each Q2D $L_p$ computation in MLSH substantially slower than that in U-HNSW.
    	Second, MLSH is built upon QALSH (Query Aware LSH~\cite{p_qalsh}), which is designed for external-memory operations and thus aims to minimize the number of sequential I/Os. As an inherent trade-off, its CPU computation time, other than that for distance computations, can be extremely large, as observed in~\cite{idec}.
    	
    	To make the comparison implementation-agnostic and extremely favorable to MLSH, we report, for MLSH, the \emph{idealized} query time that accounts only for query-to-data (Q2D) $L_p$ distance computations, calculated as $N_p T_p$, where $N_p$ is the number of Q2D $L_p$ distance computations performed by MLSH query processing (as reported by the MLSH program), and $T_p$ is the same average per 
	Q2D distance computation time as achieved in U-HNSW (using SIMD-optimized $L_p$ implementation).
    	This idealized estimate accounts only for the Q2D $L_p$ distance computations in MLSH and excludes all other expensive CPU computations.
    	As we will elaborate next, even using such an extremely favorable idealized estimate for MLSH, U-HNSW still achieves significantly shorter query times while maintaining comparable or better recall.

\subsection{Comparison against MLSH}\label{subsec:comp_mlsh}

In this subsection, we compare U-HNSW against MLSH, the state-of-the-art solution to the ANNS-U-$L_p$ problem. Since MLSH can solve this problem only for $p \leq 1$,  we restrict $p$ to values between $0.5$ and $1$. 
Accordingly, we report only the size of $G_1$ (the graph index built on the $L_1$ metric) as the index size of U-HNSW, because the graph index $G_2$ is not used (since $p \leq 1$) in this experiment.
Following~\cite{mlsh},  the $p$ value in the query tuple $(\vec{q}, p)$ is uniformly randomly selected from the following set: $\{0.5,0.6,0.7,0.8,0.9\}$.

Table~\ref{tab:eval_mlsh} clearly shows that U-HNSW significantly outperforms the idealized MLSH, achieving up to 15.0 times shorter query times while providing higher recall and requiring smaller index sizes across all six datasets. Specifically, U-HNSW achieves between 4.4 and 15.0 times shorter query times than the idealized MLSH on all datasets.

\begin{table}[!htb]
  \centering
  \setlength{\tabcolsep}{3pt}
  \renewcommand{\arraystretch}{1.0}
  \captionsetup[subtable]{skip=4pt} 
    \begin{tabular}{l ccc ccc ccc}
      \toprule
      \multirow{2}{*}{Dataset}
        & \multicolumn{2}{c}{Recall}
        & \multicolumn{2}{c}{Query Time (ms)}
        & \multicolumn{2}{c}{Index Size (MB)} \\
      \cmidrule(lr){2-3}\cmidrule(lr){4-5}\cmidrule(lr){6-7}
        & U-HNSW & MLSH & U-HNSW & MLSH & U-HNSW & MLSH \\
      \midrule
      Sun   & \textbf{0.969} & 0.942 & \textbf{1.75} & 13.75 & \textbf{21.3} & 119.5 \\
      Trevi & \textbf{0.976} & 0.951 & \textbf{8.72} & 131.22 & \textbf{26.0} & 151.2 \\
      GIST  & \textbf{0.939} & 0.917 & \textbf{9.58} & 66.83 & \textbf{260.2} & 1525.9 \\
      Deep  & \textbf{0.951} & 0.943 & \textbf{6.34} & 29.39 & \textbf{260.2} & 1525.9 \\
      GloVe & \textbf{0.926} & 0.921 & \textbf{6.01} & 26.48 & \textbf{309.4} & 1818.4 \\
      SIFT  & \textbf{0.953} & 0.942 & \textbf{1.71} & 9.12 & \textbf{985.0} & 3051.8 \\
      \bottomrule
    \end{tabular}
  \caption{Evaluation results of query times (in milliseconds), recall and index sizes (in megabytes).  Numbers in boldface are the best in each group.}
  \label{tab:eval_mlsh}
\end{table}

\subsection{Comparison with the Original HNSW}

\begin{figure}[t]
  \centering
  
  \vspace{-0.5em}
  \begin{minipage}{\linewidth}
    \centering
    \begin{tabular}{@{}c@{\hspace{0.8cm}}c@{}}
      \textcolor{red}{\rule[0.8ex]{0.9cm}{1.5pt}}
      U-HNSW &
      \textcolor{green!60!black}{\rule[0.8ex]{0.9cm}{1.5pt}}
      HNSW
    \end{tabular}
  \end{minipage}
  
  \vspace{0.5em}
  
  \begin{minipage}{\columnwidth}
    \centering
    
    \subcaptionbox{Sun\label{fig:c}}
    {\includegraphics[width=0.33\textwidth]{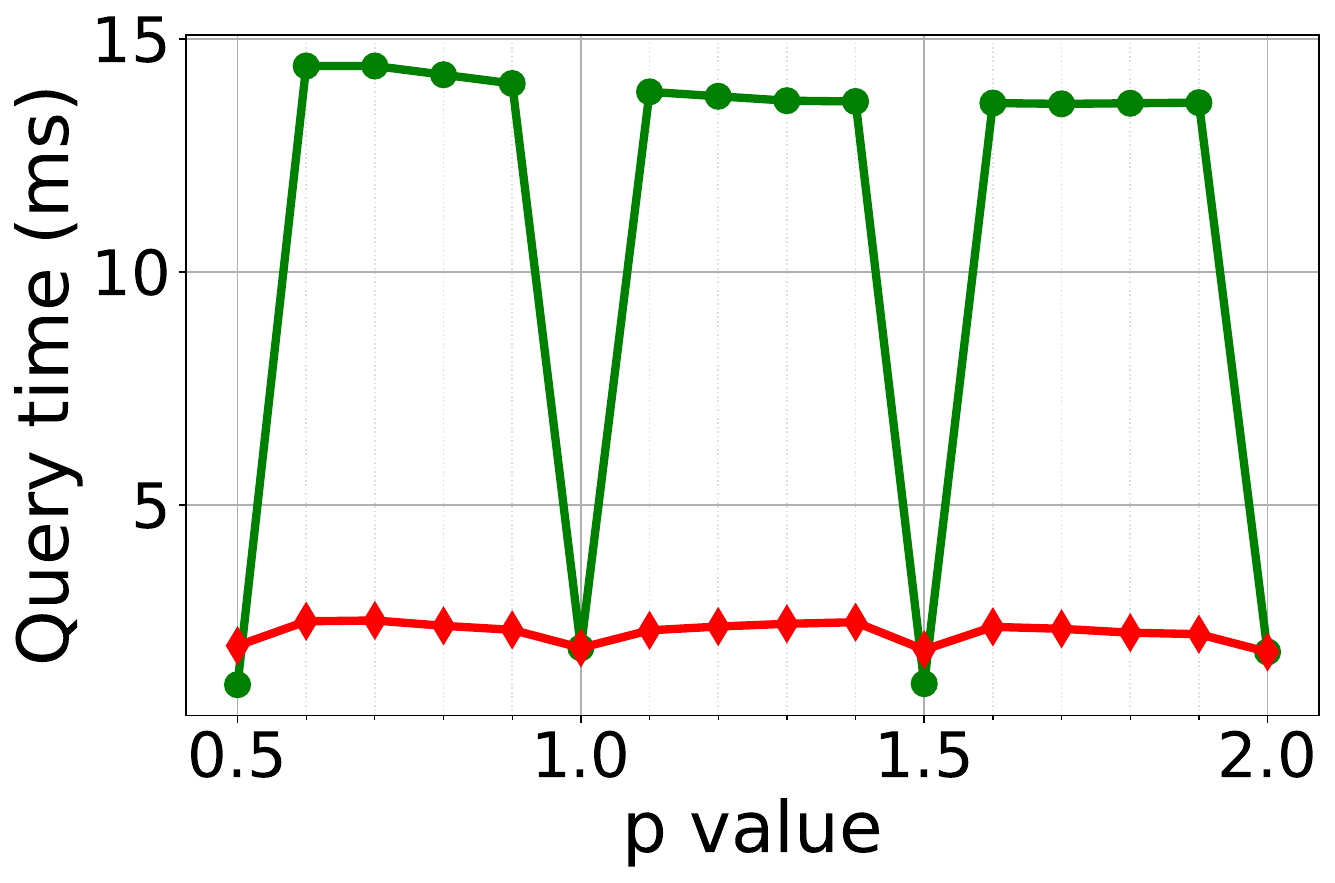}}\hfill
    \subcaptionbox{Trevi\label{fig:d}}
    {\includegraphics[width=0.33\textwidth]{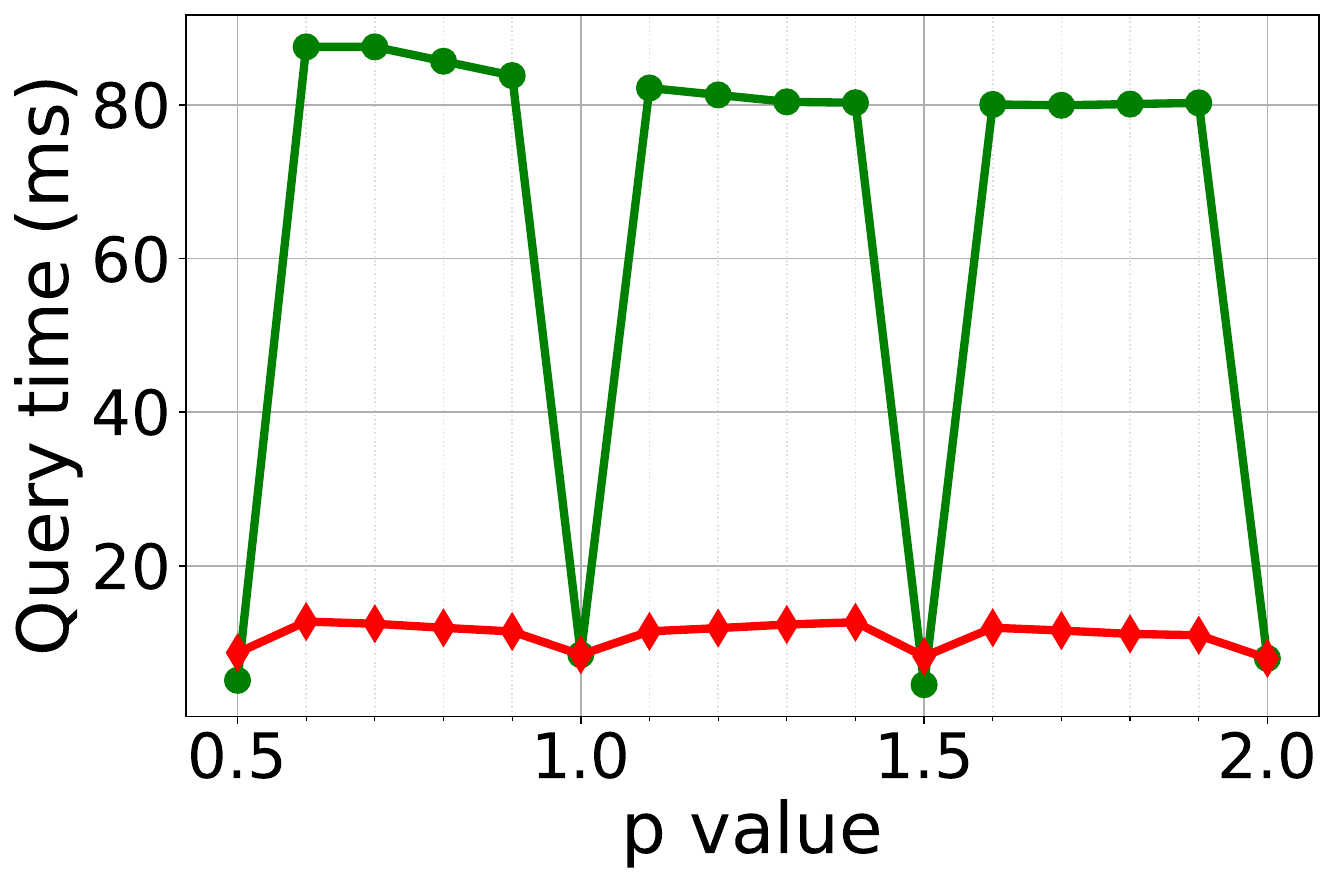}}\hfill
        \subcaptionbox{GIST\label{fig:e}}
    {\includegraphics[width=0.33\textwidth]{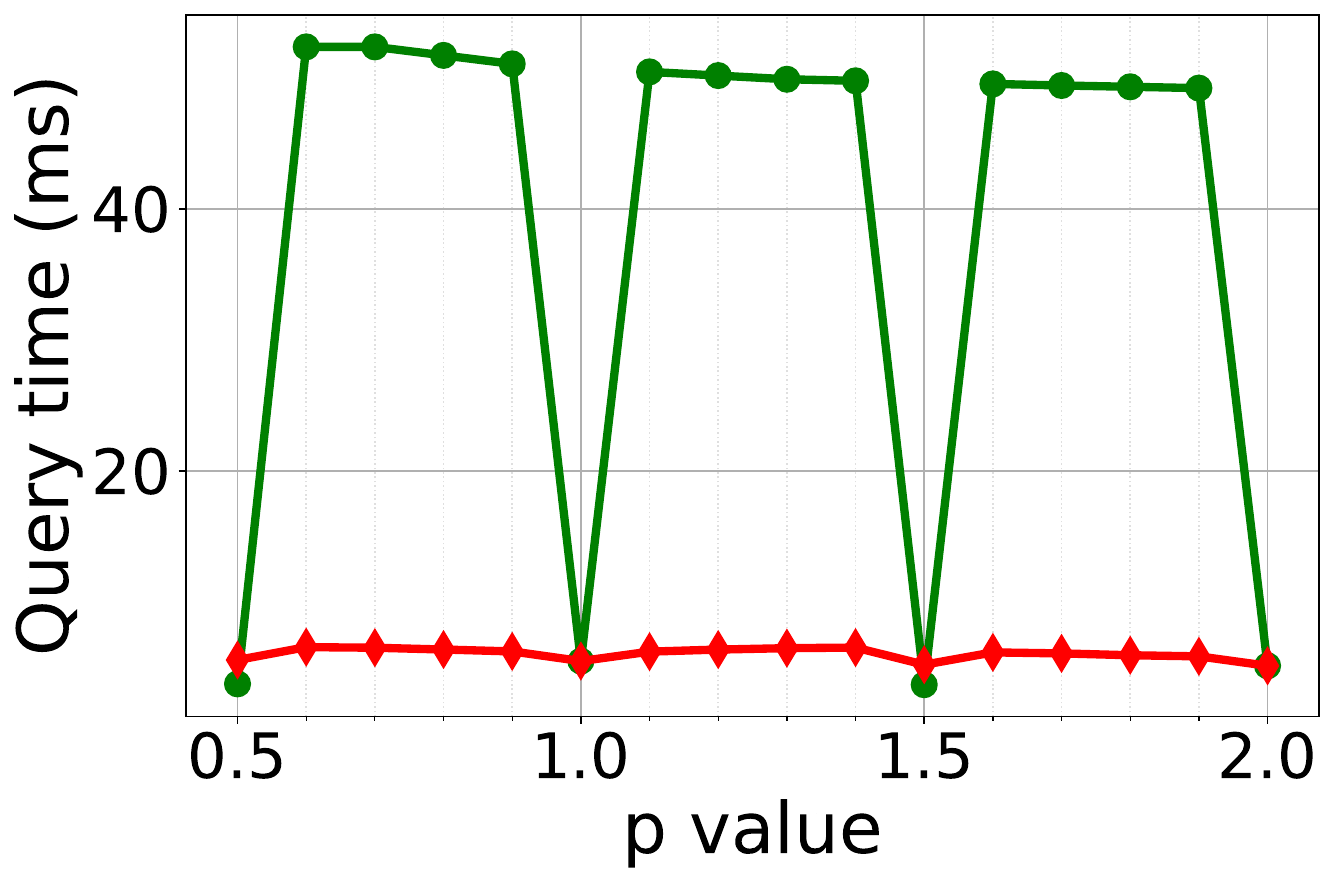}}\hfill
    

    \subcaptionbox{Deep\label{fig:f}}
    {\includegraphics[width=0.33\textwidth]{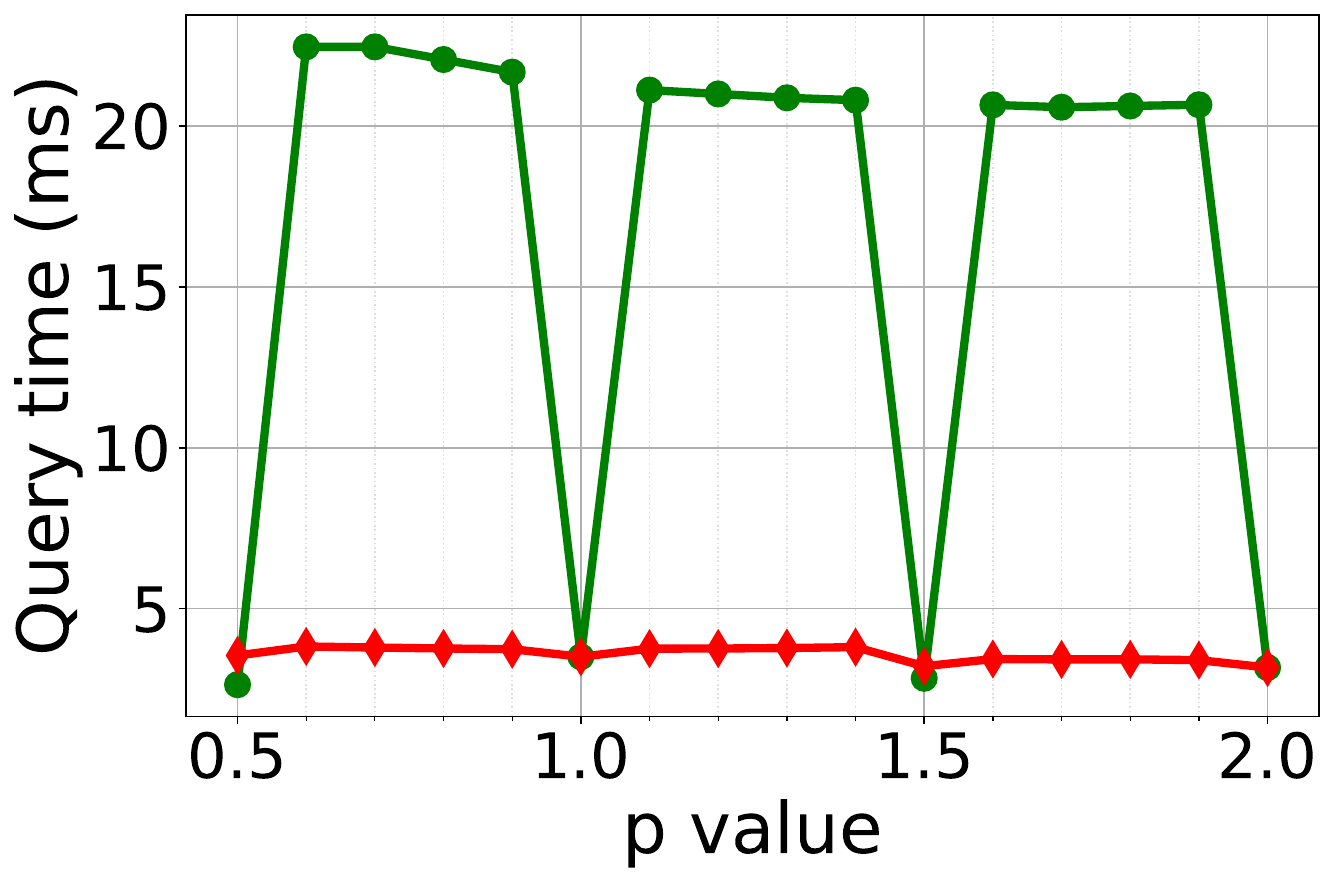}}\hfill
    \subcaptionbox{GloVe\label{fig:g}}
    {\includegraphics[width=0.33\textwidth]{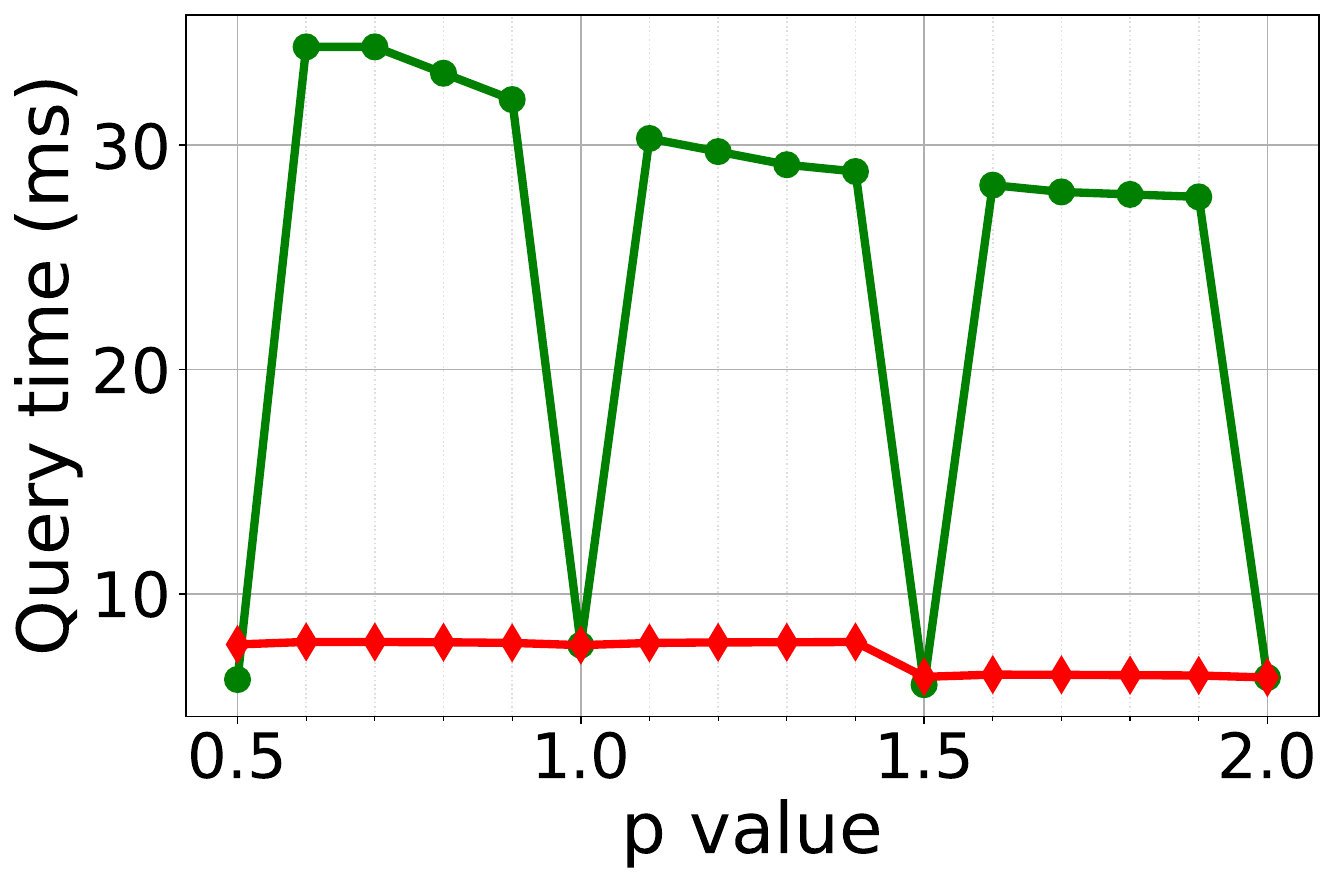}}\hfill
    \subcaptionbox{SIFT\label{fig:h}}
    {\includegraphics[width=0.33\textwidth]{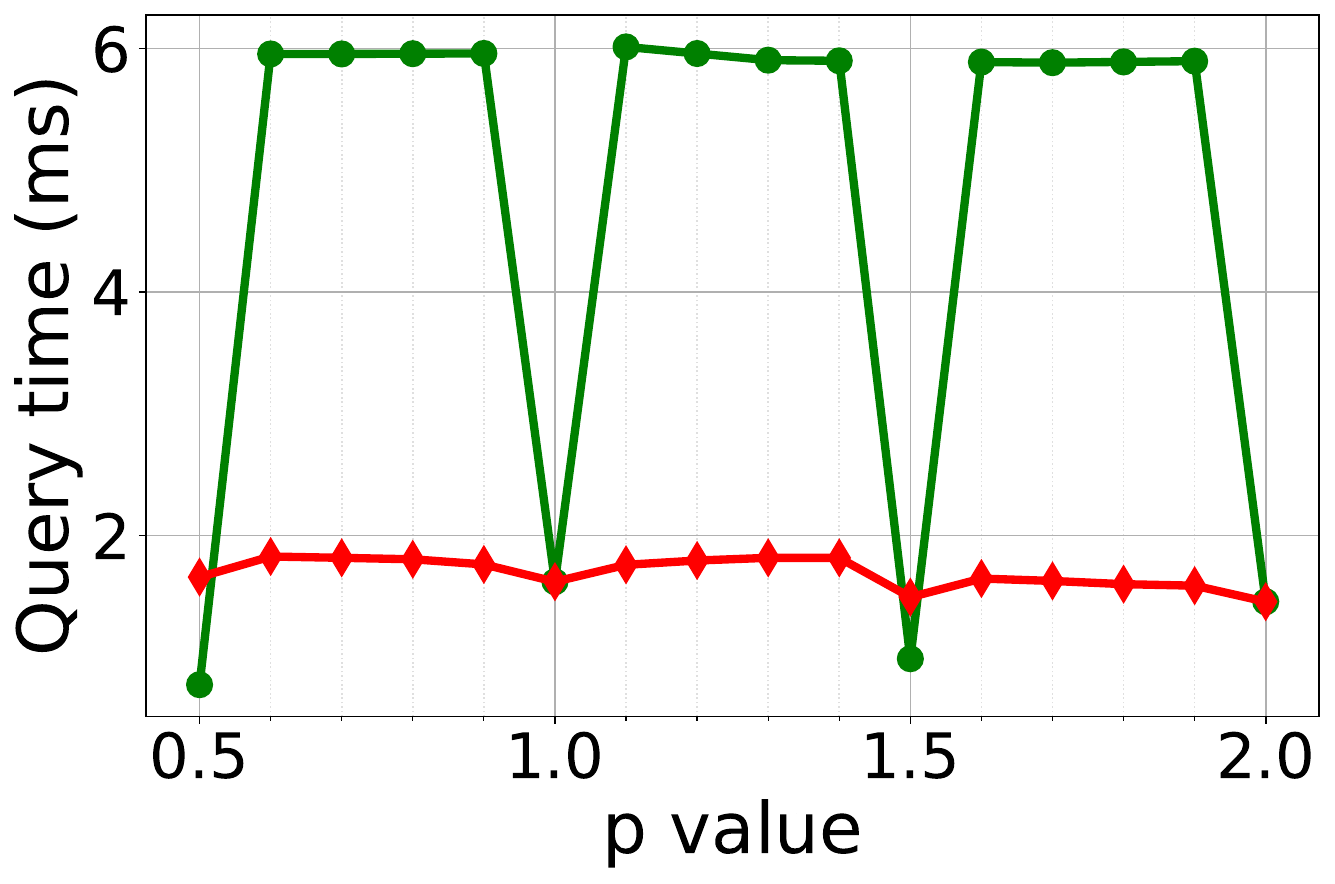}}
  \end{minipage}
  
  \vspace{0.5em}
  \caption{The query times (in milliseconds) of U-HNSW versus HNSW with different $p$ values across all datasets. Both methods achieve similar recall value and thus omitted. U-HNSW is faster than HNSW on ANNS-$L_p$ with a fixed $p$ value, except for $p=0.5$ or $1.5$.}
  \label{fig:hnsw_8figs}
\end{figure}
In this subsection, we compare U-HNSW's query performance with the original HNSW on ANNS-$L_p$ with a fixed $p$ value. The goal of this subsection is to demonstrate that our U-HNSW can achieve better query performance than the original HNSW even on the ANNS-$L_p$ problem, except for a few special $p$ values. As for the index size, U-HNSW's index size is twice that of the original HNSW because U-HNSW uses two HNSW graph indexes ($G_1$ and $G_2$). 
The evaluated $p$ values are $0.5,0.6,0.7,\cdots,2.0$.  We report the average query performance of all query points for each $p$ value.

Figure~\ref{fig:hnsw_8figs} shows the query times of U-HNSW versus those of the original HNSW for different $p$ values when both schemes achieve similar recall values higher than 0.9. 
As shown in Figure~\ref{fig:hnsw_8figs}, query times of U-HNSW are 3.3 to 8.5 times shorter than those of the original HNSW on all datasets across a wide range of $p$ values, except for $p=0.5$ and $1.5$.
At these two values, the query times of HNSW are 1.1 to 2.1 times shorter than those of U-HNSW on all datasets except GloVe, where their query performances are comparable. This is because for $p=0.5$ and $1.5$, the $L_{0.5}$ and $L_{1.5}$ distances can be efficiently computed using SIMD instructions as described in Section~\ref{subsec:anns_lp_time}, making their computation cost close to that of the base metric ($L_1$ or $L_2$).  In such a case, on the one hand, the main advantage of U-HNSW, which is to significantly reduce the number of (presumably expensive)
$L_p$ distance computations, becomes blunted.
On the other hand, U-HNSW still has to perform additional base-metric distance computations during the candidate generation step as described in Section~\ref{subsec:param}.
As a result, the overall query time of U-HNSW becomes longer than that of HNSW under these special $p$ values.
For general $L_p$ metrics that do not benefit from the SIMD instructions, however, U-HNSW consistently delivers faster query performance than the original HNSW for ANNS-$L_p$ queries.

\subsection{Ablation Study}\label{subsec:ablation}

In this subsection, we conduct a series of ablation studies to investigate the effectiveness of U-HNSW's design choices and the sensitivity of its query performance to its parameters. Specifically, we (1) demonstrate the necessity of the candidate verification step, (2) study the QPS--recall trade-off as the parameters $t$ and $\tau$ vary, and (3) compare the index construction time of U-HNSW against MLSH.

\subsubsection{Effectiveness of Candidate Verification.}\label{subsubsec:ablation_cv}
To demonstrate the necessity and effectiveness of the candidate verification step (Section~\ref{subsec:algodesc}), we conduct an ablation study at two representative $p$ values: $p=0.51$ (the most demanding general-$L_p$ case, with the largest mismatch to the base metric $L_1$) and $p=0.9$ (an easier case, close to the base metric). We use $p=0.51$ instead of $p=0.5$ because $p=0.5$ benefits from SIMD acceleration (Section~\ref{subsec:anns_lp_time}), making its verification time artificially short; $p=0.51$ is equally demanding but gives a fairer time breakdown. Results are shown in Table~\ref{tab:ablation}.

\begin{table*}[t]
	\centering
	\caption{Ablation study of U-HNSW's candidate verification. ``Recall after initial filtering'' is the recall of directly returning the first $K$ points of $C$ sorted by $L_1$ without re-ranking; ``Recall after re-ranking'' is the recall after candidate verification. Candidate generation time is the same for both $p$ values. ``Verification without early stop'' is the hypothetical candidate verification time if all $t=300$ candidates were checked without early termination.}
	\label{tab:ablation}
	\footnotesize
	\setlength{\tabcolsep}{6pt}
	\renewcommand{\arraystretch}{1.2}
	\begin{tabular}{l rr rr c rr rr}
		\toprule
		& \multicolumn{2}{c}{Recall after initial filtering}
		& \multicolumn{2}{c}{Recall after re-ranking}
		& \multicolumn{1}{c}{\makecell{Candidate\\Generation (ms)}}
		& \multicolumn{2}{c}{Candidate Verification (ms)}
		& \multicolumn{2}{c}{Verification without early stop (ms)} \\
		\cmidrule(lr){2-3}\cmidrule(lr){4-5}\cmidrule(lr){6-6}
		\cmidrule(lr){7-8}\cmidrule(lr){9-10}
		Dataset
		& $p{=}0.51$ & $p{=}0.9$
		& $p{=}0.51$ & $p{=}0.9$
		&
		& $p{=}0.51$ & $p{=}0.9$
		& $p{=}0.51$ & $p{=}0.9$ \\
		\midrule
		Sun   & 0.751 & 0.952 & 0.976 & 1.000 & 1.93 & 0.65 & 0.40 & 1.24 & 1.18 \\
		Trevi & 0.789 & 0.958 & 0.979 & 0.998 & 8.42 & 4.77 & 3.23 & 9.43 & 9.53 \\
		GIST  & 0.747 & 0.950 & 0.951 & 0.975 & 5.47 & 1.20 & 0.77 & 2.33 & 2.25 \\
		Deep  & 0.776 & 0.958 & 0.924 & 0.960 & 3.50 & 0.37 & 0.28 & 0.71 & 0.80 \\
		GloVe & 0.659 & 0.937 & 0.907 & 0.961 & 7.71 & 0.16 & 0.10 & 0.28 & 0.28 \\
		SIFT  & 0.632 & 0.938 & 0.932 & 0.996 & 1.62 & 0.25 & 0.15 & 0.43 & 0.41 \\
		\bottomrule
	\end{tabular}
\end{table*}

Table~\ref{tab:ablation} shows three observations.

\noindent\textbf{(1)} Recall after initial filtering is already 0.94--0.96 at $p=0.9$, suggesting that for easy cases (where the $L_p$ metric is close to the base metric), simply returning the first $K$ points of $C$ sorted by the base metric $L_1$ may seem acceptable. However, at $p=0.51$, recall after initial filtering drops to only 0.63--0.79, far below the target recall of 0.9. Since ANNS-U-$L_p$ must handle arbitrary $p$ values per query, relying on initial filtering alone is insufficient.

\noindent\textbf{(2)} After re-ranking with exact $L_p$ distances, the final recall rises to 0.90--0.98 at $p=0.51$ and 0.96--1.00 at $p=0.9$, meeting the target recall of 0.9. Since ANNS-U-$L_p$ queries can arrive with any $p$ value, candidate verification is essential to guarantee acceptable recall across both easy and demanding cases.

\noindent\textbf{(3)} The early-termination strategy is crucial for keeping candidate verification from dominating the total query time. On the Trevi dataset (a relatively small dataset, see Table~\ref{tab:dataset-stats}), without early termination, candidate verification would take 9.43~ms at $p=0.51$ and 9.53~ms at $p=0.9$, increasing total query time from 13.19~ms to 17.85~ms ($1.35\times$) and from 11.65~ms to 17.95~ms ($1.54\times$), respectively. With early termination, candidate verification is reduced to 4.77~ms and 3.23~ms, staying well below the candidate generation time of 8.42~ms. Therefore, early termination is particularly valuable on small datasets where, without it, candidate verification time would match or exceed candidate generation time.

\subsubsection{Sensitivity to Parameters $t$ and $\tau$.}\label{subsubsec:ablation_t_tau}

We further study how the parameters $t$ and $\tau$ affect U-HNSW's QPS--recall trade-off on GIST and SIFT. We use $\mathit{efSearch} = 500$ to accommodate the varying $t$ values (from 100 to 500), and sample $p$ uniformly from $\{0.5, 0.6, \ldots, 2.0\}$ per query to reflect the true ANNS-U-$L_p$ setting.

\begin{figure}[t]
	\centering
	\begin{minipage}{\linewidth}
		\centering
		\scriptsize
		\setlength{\tabcolsep}{6pt}
		\begin{tabular}{@{}cc@{}}
			\textcolor{green!60!black}{\rule[0.8ex]{0.8cm}{1.2pt}}~GIST&
			\textcolor{red}{\rule[0.8ex]{0.8cm}{1.2pt}}~SIFT
		\end{tabular}
	\end{minipage}\\[2pt]
	\begin{subfigure}[b]{0.235\textwidth}
		\centering
		\includegraphics[width=\textwidth]{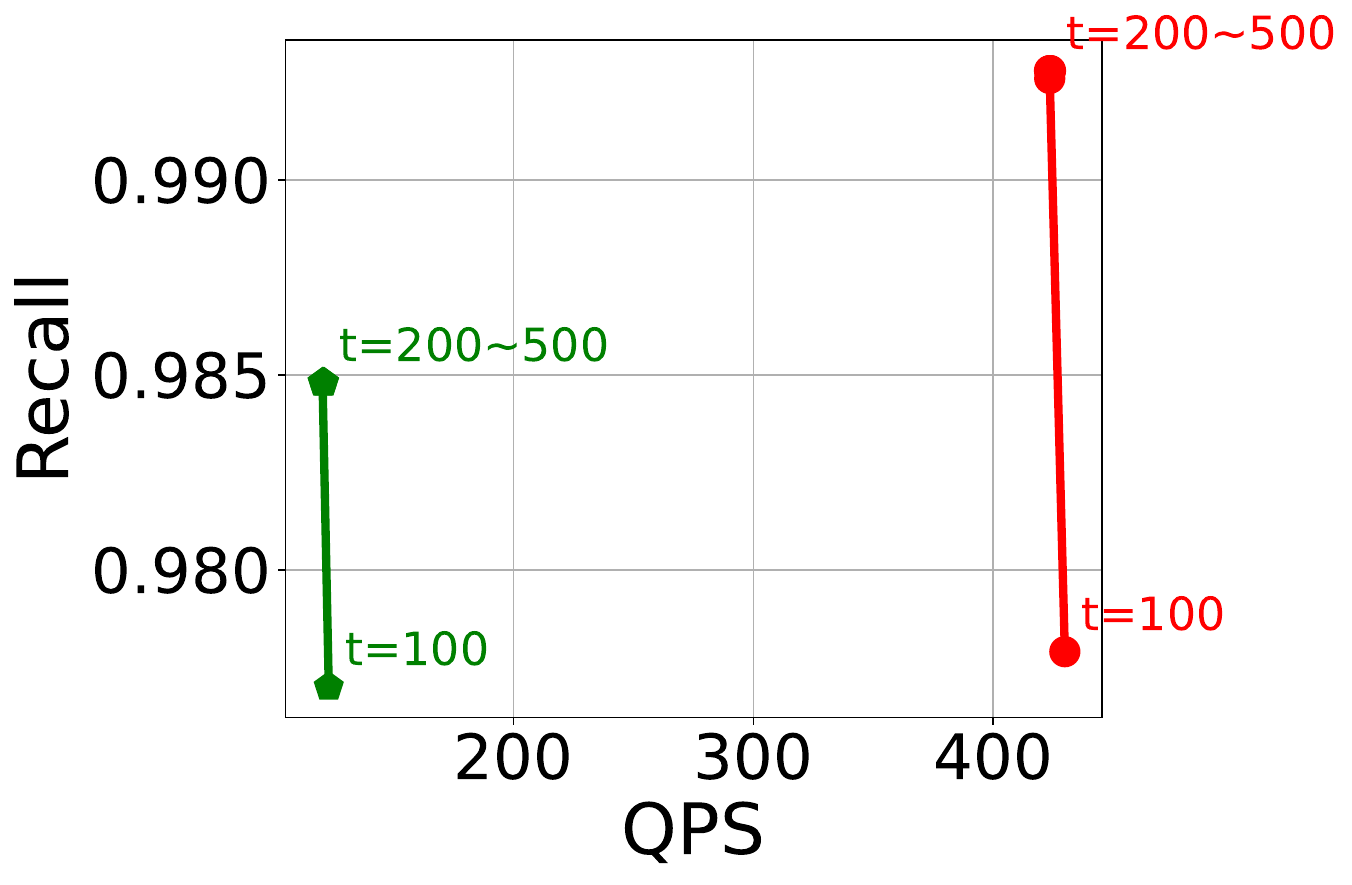}
		\caption{QPS--Recall under different $t$.}
		\label{fig:qps_recall_t}
	\end{subfigure}
	\hfill
	\begin{subfigure}[b]{0.235\textwidth}
		\centering
		\includegraphics[width=\textwidth]{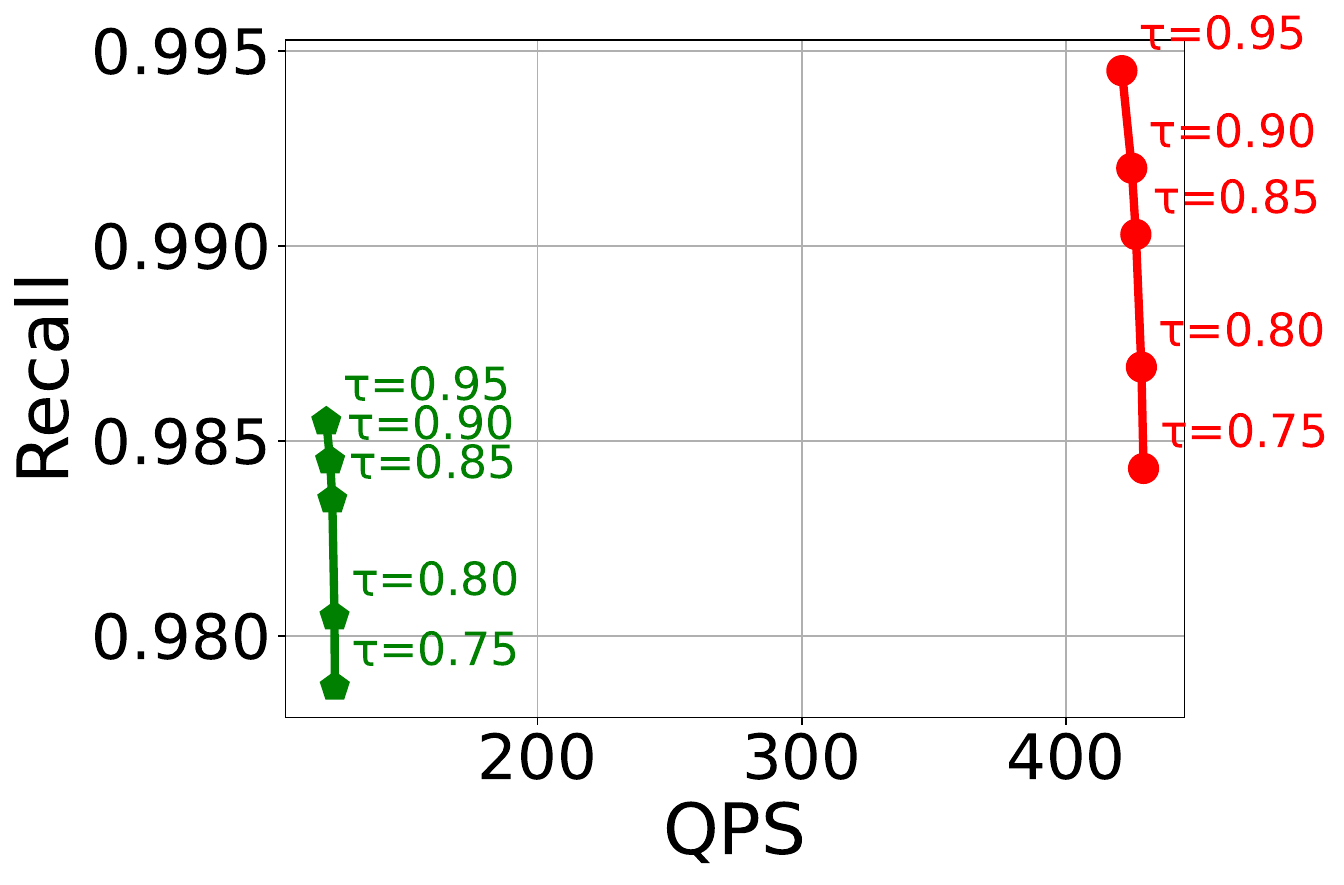}
		\caption{QPS--Recall under different $\tau$.}
		\label{fig:qps_recall_tau}
	\end{subfigure}
	\caption{QPS--recall curves of U-HNSW on GIST and SIFT for $p \in [0.5, 2.0]$ (uniformly sampled per query). When $t \geq 200$, U-HNSW achieves similar query performance, and $\tau \in [0.90, 0.95]$ provides the best trade-off between QPS and recall.}
	\label{fig:qps_recall}
\end{figure}

Figure~\ref{fig:qps_recall_t} shows that varying $t$ from 200 to 500 yields similar QPS and recall, confirming that $t = 300$ is a conservative choice that achieves the near-optimal QPS--recall trade-off.

Figure~\ref{fig:qps_recall_tau} shows the speed-accuracy trade-off when varying $\tau$: decreasing $\tau$ from 0.95 to 0.75 increases QPS while reducing recall. Since $p$ is uniformly sampled from $\{0.5, 0.6, \ldots, 2.0\}$ per query, many queries have $p$ values close to the base metric, making candidate verification easier and recall remains high even at smaller $\tau$. However, our evaluation also includes ANNS-$L_p$ queries with $p = 0.5$, the most demanding case as the mismatch between the base metric $L_1$ and the target $L_{0.5}$ is the largest. To reliably achieve recall above 0.9 even under this demanding case, we set $\tau = 0.92$ as the default, which provides a sufficient safety margin across all datasets and $p \in [0.5, 2]$.

\subsubsection{Index Construction Time.}\label{subsubsec:ablation_build}

\begin{table}[t]
	\centering
	\setlength{\tabcolsep}{4pt}
	\caption{Index construction time (in seconds) for U-HNSW (total time of building $G_1$ and $G_2$) and MLSH (CPU only, excluding disk I/O).}
	\label{tab:build_time}
	\small
	\begin{tabular}{lrrrrrr}
		\toprule
		Dataset & Sun & Trevi & GIST & Deep & GloVe & SIFT \\
		\midrule
		U-HNSW & 205 & 1{,}225 & 8{,}200 & 3{,}175 & 2{,}250 & 4{,}050 \\
		MLSH   &  20 &    165  &    422  &    143  &     96  &    186  \\
		\bottomrule
	\end{tabular}
\end{table}

Table~\ref{tab:build_time} compares the index construction time of U-HNSW (the total time of building $G_1$ and $G_2$) with that of MLSH (CPU only, excluding disk I/O). MLSH's CPU build time is significantly shorter than U-HNSW's. However, U-HNSW's construction time remains competitive for the following two reasons.

First, MLSH is built on QALSH~\cite{p_qalsh}, an external-memory index that writes paged data and B$^+$-tree files to disk during construction, incurring heavy I/O overhead. When this disk I/O cost is included, the gap largely disappears: on the Sun dataset, MLSH's total construction time reaches 200~seconds, comparable to U-HNSW's 207~seconds (which includes U-HNSW's own disk I/O time). Furthermore, as described in~\cite{ann_benchmark}, when sufficient computing resources are available for offline index construction, graph-based methods such as HNSW are recommended as the best choice for ANNS due to their outstanding search performance---despite having longer construction times than LSH-based methods. This is because index construction is a one-time offline cost: once built, the graph can be reused across all subsequent queries~\cite{cagra}.

Second, U-HNSW is far more construction-efficient than the naive alternative of building a separate HNSW index for each possible $p$ value. For example, on GIST, building a single HNSW index for a general $p$ would take more than 10 hours, and building indexes for even a modest number (say, 10) of $p$ values would be practically unacceptable. By contrast, U-HNSW requires building only $G_1$ (on the $L_1$ metric) and $G_2$ (on the $L_2$ metric), whose construction times are much shorter due to the computation efficiency of $L_1$ and $L_2$.

\subsection{U-HNSW Extended Version (U-HNSW-E)}\label{subsec:uhnsw_e}

In our default configuration, U-HNSW targets the commonly evaluated range $p \in [0.5, 2]$ used in prior ANNS-U-$L_p$ works~\cite{lazylsh,mlsh}. If the application range of $p$ is shifted toward smaller values (e.g., $p \in [0.2, 1]$), U-HNSW can be further extended as follows. First, instead of building and using graph indexes $G_1$ and $G_2$, here we build and use $G_{0.5}$ (an HNSW graph index built on the $L_{0.5}$ metric) and $G_1$ (on the $L_1$ metric). Second, we modify the cutoff in Algorithm~\ref{alg:lpquery} from $p \leq 1.4$ to $p \leq 0.6$ (and replace $G_1$ and $G_2$ in the algorithm with $G_{0.5}$ and $G_1$, respectively). We refer to this extended scheme as U-HNSW-E.

\subsubsection{Comparison with MLSH on the Shifted $p$-range.}
We provide the evaluation result of U-HNSW-E versus MLSH on the shifted $p$-range $[0.2,0.7]$ used in~\cite{mlsh}, on two representative datasets, Deep and SIFT, which are also used in the MLSH paper~\cite{mlsh}. As shown in Table~\ref{tab:eval_mlsh_extended}, U-HNSW-E remains faster than the idealized MLSH in this shifted $p$-range, by a factor of 4.8 on Deep and 5.8 on SIFT, while still maintaining higher recall and a smaller index size across the extended $p$-range.

\begin{table}[!htb]
	\centering
	\setlength{\tabcolsep}{3pt}
	\renewcommand{\arraystretch}{1.0}
		\caption{Evaluation results on ANNS-U-$L_p$ for $0.2 \le p \le 0.7$ (extended range). Numbers in boldface are the best in each group.}
	\resizebox{\columnwidth}{!}{%
	\begin{tabular}{l ccc ccc ccc}
		\toprule
		\multirow{2}{*}{Dataset}
		& \multicolumn{2}{c}{Recall}
		& \multicolumn{2}{c}{Query Time (ms)}
		& \multicolumn{2}{c}{Index Size (MB)} \\
		\cmidrule(lr){2-3}\cmidrule(lr){4-5}\cmidrule(lr){6-7}
		& U-HNSW-E & MLSH & U-HNSW-E & MLSH & U-HNSW-E & MLSH \\
		\midrule
		Deep & \textbf{0.944} & 0.921 & \textbf{7.15}  & 34.37 & \textbf{520.4}  & 1525.9 \\
		SIFT & \textbf{0.948} & 0.937 & \textbf{1.95}  & 11.24 & \textbf{1970.0} & 3051.8 \\
		\bottomrule
	\end{tabular}}
	\label{tab:eval_mlsh_extended}
\end{table}

\subsubsection{Comparison with HNSW on the Shifted $p$-range.}
\begin{figure}[t]
	\centering
	\begin{minipage}{\linewidth}
		\centering
		\scriptsize
		\setlength{\tabcolsep}{6pt}
		\begin{tabular}{@{}cc@{}}
			\textcolor{red}{\rule[0.8ex]{0.9cm}{1.5pt}}~U-HNSW-E &
			\textcolor{green!60!black}{\rule[0.8ex]{0.9cm}{1.5pt}}~HNSW
		\end{tabular}
	\end{minipage}\\[2pt]
	\begin{subfigure}[b]{0.235\textwidth}
		\centering
		\includegraphics[width=\textwidth]{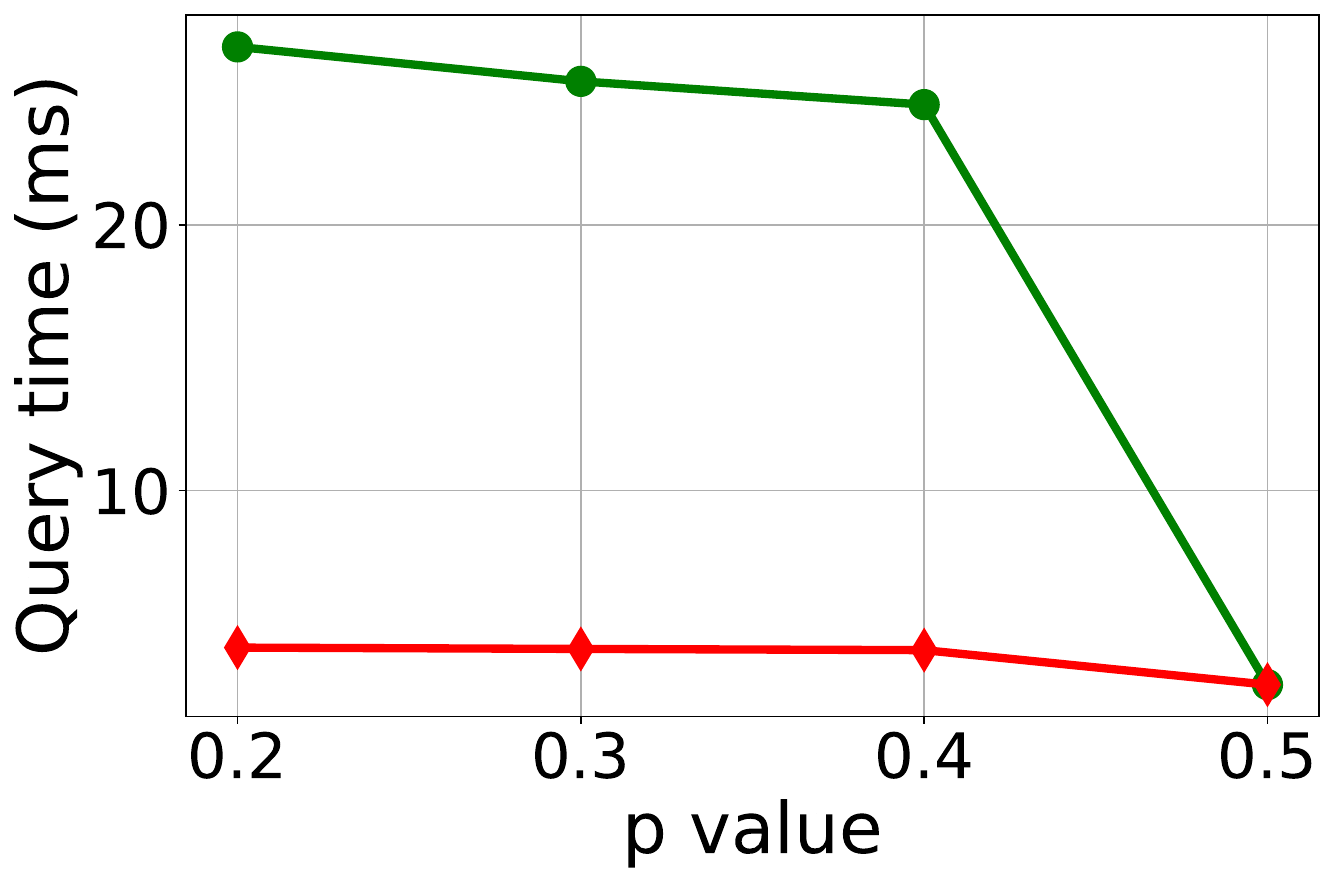}
		\caption{Deep}
		\label{fig:extend_deep}
	\end{subfigure}
	\hfill
	\begin{subfigure}[b]{0.235\textwidth}
		\centering
		\includegraphics[width=\textwidth]{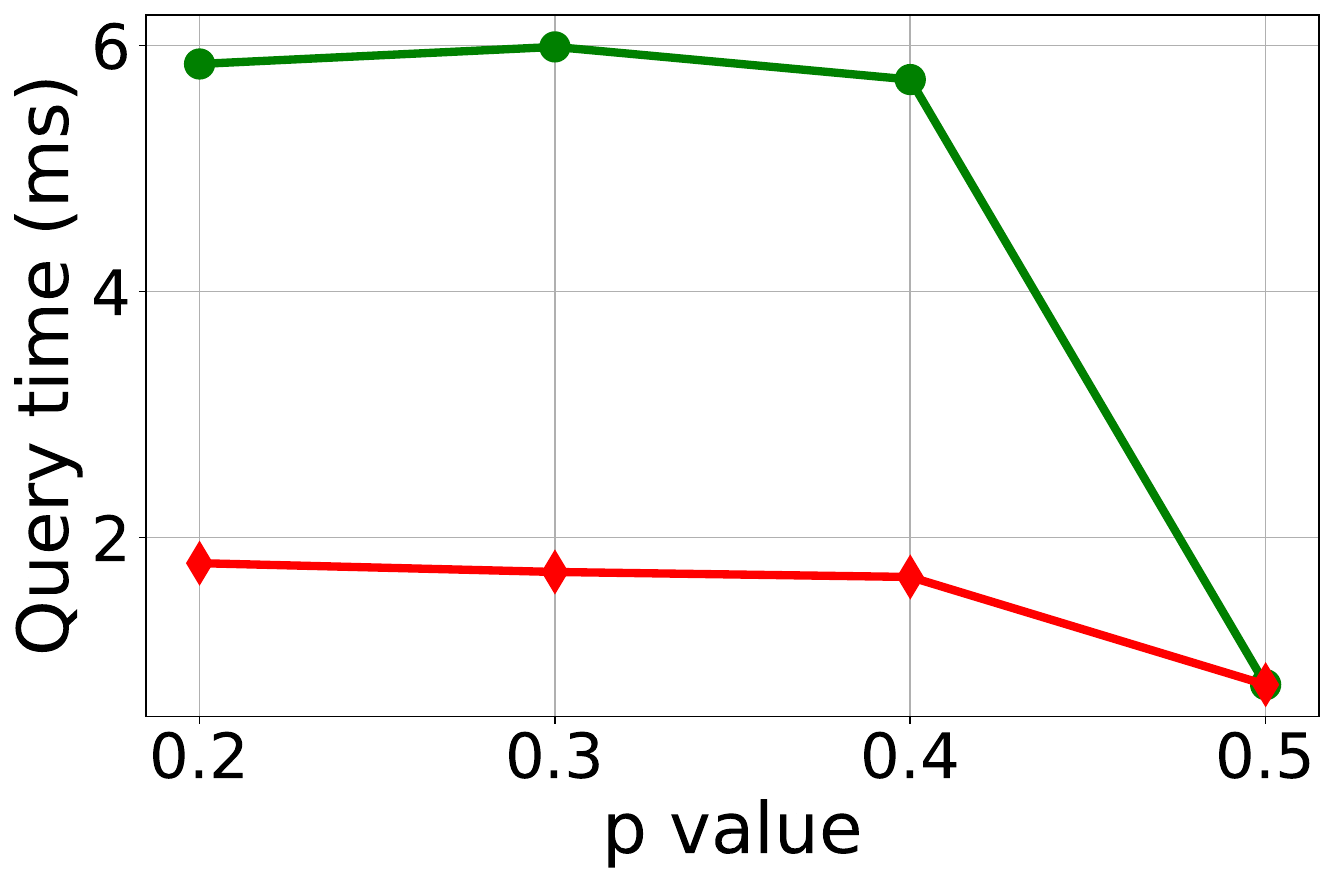}
		\caption{SIFT}
		\label{fig:extend_sift}
	\end{subfigure}
	\caption{Query times (in milliseconds) of U-HNSW-E versus HNSW with different $p$ values on Deep and SIFT. U-HNSW-E achieves higher or similar recall to HNSW.}
	\label{fig:uhnsw_hnsw_extend_p}
\end{figure}

We further compare U-HNSW-E against the original HNSW on the same two representative datasets, Deep and SIFT, for $p \in [0.2, 0.5]$, as shown in Figure~\ref{fig:uhnsw_hnsw_extend_p}. The figure reports the query times of U-HNSW-E and HNSW for different $p$ values, where U-HNSW-E achieves higher or similar recall to HNSW. As shown in Figure~\ref{fig:uhnsw_hnsw_extend_p}, U-HNSW-E consistently outperforms HNSW except for $p = 0.5$. At $p = 0.5$, the performance of U-HNSW-E matches that of HNSW, because U-HNSW-E directly queries $G_{0.5}$, which is itself an HNSW index built on $L_{0.5}$.

\section{Conclusion}
In this paper, we propose U-HNSW, the first graph-based solution for approximate nearest neighbor search under universal $L_p$ metrics.
U-HNSW leverages two base graph indexes built on two base metrics (\(L_1\) and \(L_2\)) to generate high-recall candidates and then verifies these candidates with an early-termination strategy that substantially reduces the number of expensive $L_p$ distance computations.
Experiments on multiple real-world datasets show that  U-HNSW not only achieves up to 2670 times shorter query times than the original MLSH implementation running on a RAM disk (up to 15 times shorter than the idealized MLSH), but also outperforms the original HNSW on the ANNS-$L_p$ problem (with a fixed $p$ value), except for a few special $p$ values. 



%
%
%
%
\bibliographystyle{ACM-Reference-Format}
\bibliography{bib/graph.bib}

\end{document}